\documentclass[11 pt]{article}
\usepackage{amssymb,latexsym,amsmath,amsfonts}
\usepackage{subfigure,color}
\usepackage{graphicx, epsf}
\usepackage[active]{srcltx}
\usepackage[active]{srcltx}
\numberwithin{equation}{section}

\usepackage{amssymb,latexsym,amsmath,amsfonts}
\usepackage{graphicx, epsf, subfigure}
\usepackage{epsfig}

\begin{document}

\title{Regular and Singular Pulse and Front Solutions and Possible Isochronous Behavior in the Short-Pulse Equation: Phase-Plane, Multi-Infinite Series and Variational Approaches}

\author{G. Gambino\footnote{Department of Mathematics, University of Palermo, Italy, gaetana@math.unipa.it}$\;$
U. Tanriver\footnote{Department of Mathematics, Texas A\&M University-Texarkana, USA, utanriver@tamut.edu}$\;$
P. Guha\footnote{S.N. Bose National Centre for Basic Sciences, Kolkata, India, partha@bose.res.in}$\;$
A. Ghose Choudhury\footnote{Department of Physics, Surendranath College, Kolkata, India, aghosechoudhury@rediffmail.com}$\;$
S. Roy Choudhury\footnote{Department of Mathematics, University of Central Florida, Orlando, USA, choudhur@cs.ucf.edu}}

\maketitle

\begin{abstract}

In this paper we employ three recent analytical approaches to investigate the possible
classes of traveling wave solutions of some members of a family of so-called short-pulse equations (SPE).
A recent, novel application
of phase-plane analysis is first employed to show the existence of breaking kink wave solutions in certain parameter regimes.
Secondly, smooth traveling waves are derived using a recent technique to derive convergent multi-infinite series solutions for
the homoclinic (heteroclinic) orbits of the traveling-wave equations for the SPE equation, as well as for
its generalized version with arbitrary coefficients. These correspond to pulse (kink
or shock) solutions respectively of the original PDEs. 
\vskip 1.5cm
Unlike the majority of unaccelerated convergent series, high
accuracy is attained with relatively few terms. And finally, variational methods are employed to generate families of both regular and embedded solitary wave solutions for the SPE PDE.
The technique for obtaining the embedded solitons incorporates several recent generalizations of the usual variational technique and it is thus topical in itself. One unusual feature of the solitary waves derived here is that we are able to obtain them in analytical form (within the assumed ansatz for the trial functions). Thus, a direct error analysis is performed, showing the accuracy of the resulting solitary waves. Given the importance of solitary wave solutions in wave dynamics and information propagation in nonlinear PDEs, as well as the fact that not much is known
about solutions of the family of generalized SPE equations considered here, the results obtained are both new and timely.

\end{abstract}




\section{Introduction}\label{Sec0}

In the paper \cite{SS06}, an exact nonsingular solitary wave solution was derived for the short-pulse (SPE) equation \cite{Scha04}, an important recent alternative to the very
widely-studied nonlinear Schrodinger (NLS) equation for ultra-short light pulses in optical fibers, where the pulse spectrum is not
narrowly localized around the carrier frequency. In this paper, we investigate additional solutions of this important recent model equation by three other techniques.

Various analytical methods have been developed to construct solitary waves of physically important nonlinear partial differential equations (NLPDEs), including variational methods,
diverse series solution techniques, the extended $\tanh-$function method, Hirota's method, truncated regular and invariant Painlev\'e expansions, and various others.

Three of these techniques are applied to the SPE equation in this paper. First, novel phase-plane methods are used to consider singular solutions of the SPE equation, in particular breaking kink or front solutions. We next employ one recently developed technique to construct convergent, multi-infinite, series solutions for regular solitary waves of the SPE equation (or equivalently, homoclinic orbits of its traveling-wave equation). In addition, in an alternative approach, the variational method is employed to construct regular solitary waves of the SPE NLPDE directly,
and also attempt to construct embedded solitons of the PDE using several recent extensions of the variational approach.

The remainder of the paper is organized as follows. In Section 2, the traveling wave ODE of the SPE equation is considered.
A recently developed technique (see \cite{CG13}, \cite{RGC14}) is employed to construct convergent series solutions for its homoclinic and heteroclinic orbits, corresponding to solitary wave and front (pulse) solutions of the original SPE NLPDE. A Lagrangian for the SPE equation is developed in Section 3. Section 4 then considers the linear spectrum of the SPE equation to isolate the parameter regimes where regular solitary waves exist. A Gaussian ansatz or trial function for these solitary waves is then substituted into the Lagrangian and its Euler-Lagrange equations are solved to derive the optimum soliton or ansatz parameters in the usual way (within the functional Gaussian form of the ansatz).

\section{Singular solutions of the SPE }\label{Sec1}

In this section, we will consider regular pulse and front solutions of the SPE \eqref{SPE}
by calculating convergent, multi-infinite, series solutions for the possible homoclinic orbits of its traveling wave equation \eqref{SPE_trav}.

Let us first consider the following short pulse equation (SPE):
\begin{equation}
\label{SPE}
u_{xt}=u+\frac{1}{6}(u^3)_{xx},
\end{equation}

\noindent where $u = u(x, t)$. This was derived in \cite{Scha04} as a model
equation describing the propagation of ultra-short light pulses in silica optical fibres.
Substituting $u(x,t)=u(x+ct)=u(z)$, where $z=x+ct$ and $c$ is the wave speed, into Eq. \eqref{SPE} we obtain:
\begin{equation}\label{SPE_trav}
(u^2-2c)u_{zz}+2u(1+u_z^2)=0.
\end{equation}
Eq. \eqref{SPE_trav} is equivalent to the following $2$-dimensional system:
\begin{equation}\label{SPE_sd}
\left\{\begin{array}{ll}
\displaystyle\frac{du}{dz}=y,\\
\,\\
\displaystyle\frac{dy}{dz}=\frac{2u(1+y^2)}{2c-u^2},
\end{array}\right.
\end{equation}
which is the traveling wave system for \eqref{SPE}.

The system \eqref{SPE_sd} belongs to the first type (see \cite{LD07}, \cite{RGC14}) of singular traveling wave system as in \eqref{gen_ld}:
\begin{equation}\label{gen_ld}
\frac{d u}{d z}=y,\qquad\qquad \frac{d y}{d z}=-\frac{G'(u)y^2+F(u)}{G(u)},
\end{equation}
where $F$ and $G$ are the following smooth nonlinear functions:
\begin{equation}
\label{SPE_pos}
F(u)=-2u,\qquad G(u)=2c-u^2.
\end{equation}

When $c>0$ the function $G(u)$ has two real zeros:
\begin{equation}
\label{SPE_zero_sing}
u_{s,1}=\sqrt{2c}\quad {\rm and}\quad u_{s,2}=-\sqrt{2c}.
\end{equation}
The function $F(u)$ admits just one zero $u_r=0$ and $F'(u_r)=-2<0$, therefore the point $P\equiv (u_r, 0)\equiv (0, 0)$ is a regular equilibrium of the system \eqref{SPE_sd}.
Moreover, the second equation in \eqref{SPE_sd} is discontinuous in the straight lines $u=u_{s,1}=\sqrt{2c}$ and $u=u_{s,2}=-\sqrt{2c}$.
Being $Y=-\displaystyle\frac{F(u_{s,1})}{G'(u_{s,1})}=-\frac{F(u_{s,2})}{G'(u_{s,2})}=-1<0$, there are no singular real equilibria in the singular straight lines $u=u_{s,1}$ and $u=u_{s,2}$ (this
is due to the fact that $F(u)=G'(u)$).

\begin{figure}
\begin{center}
\subfigure[] {\epsfxsize=2.7 in \epsfbox{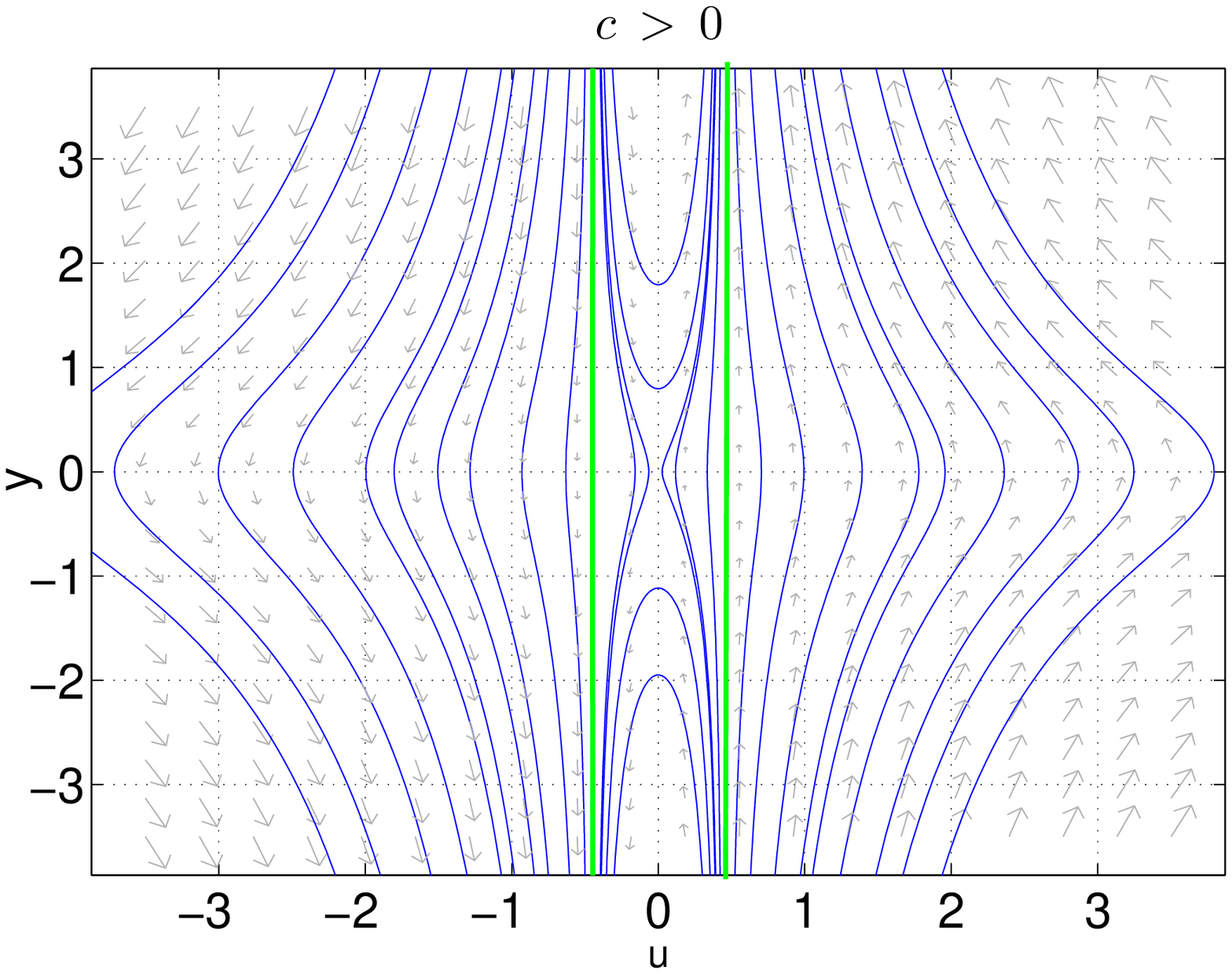}}
\subfigure[] {\epsfxsize=2.5 in \epsfbox{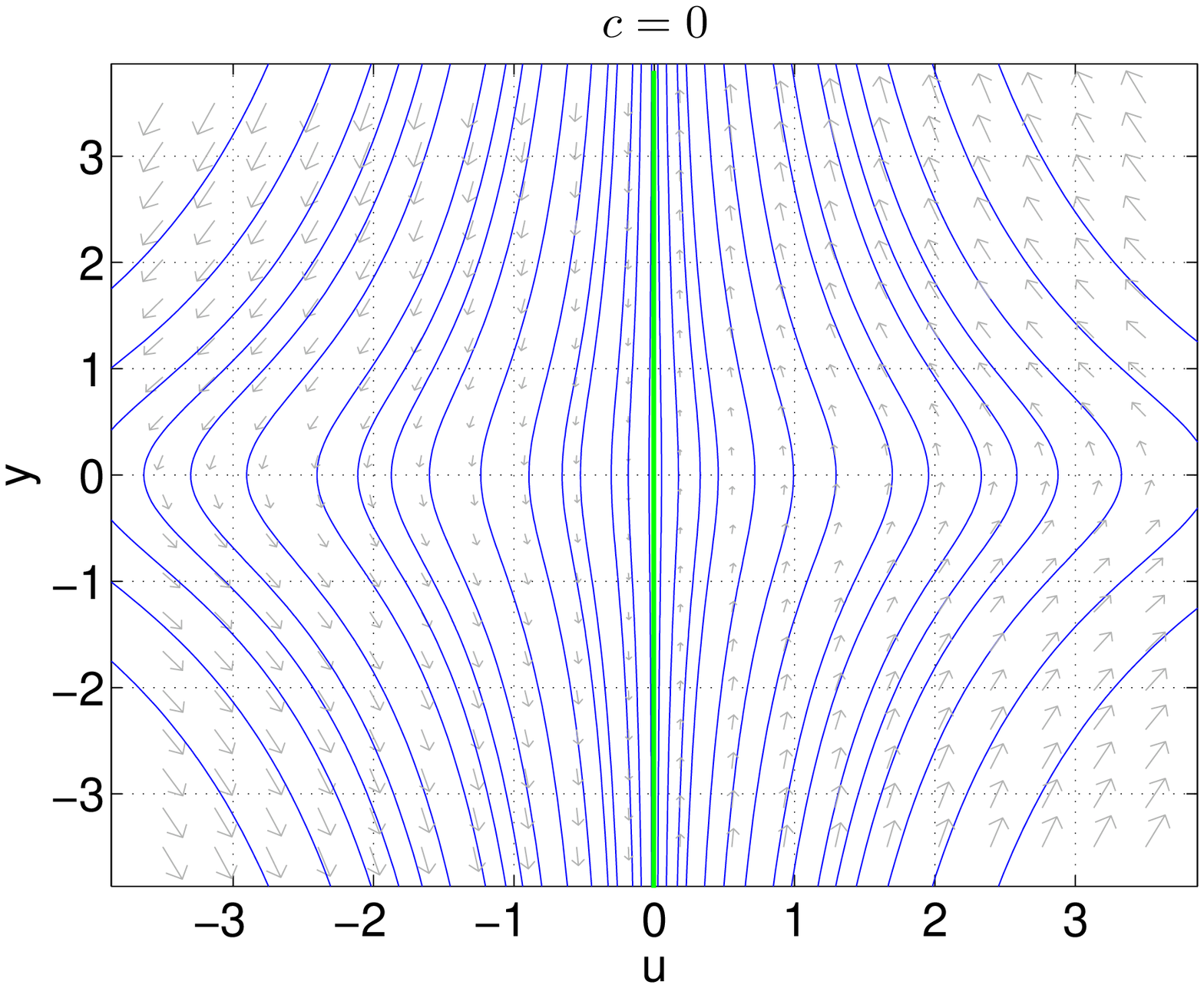}}
\subfigure[] {\epsfxsize=2.5 in \epsfbox{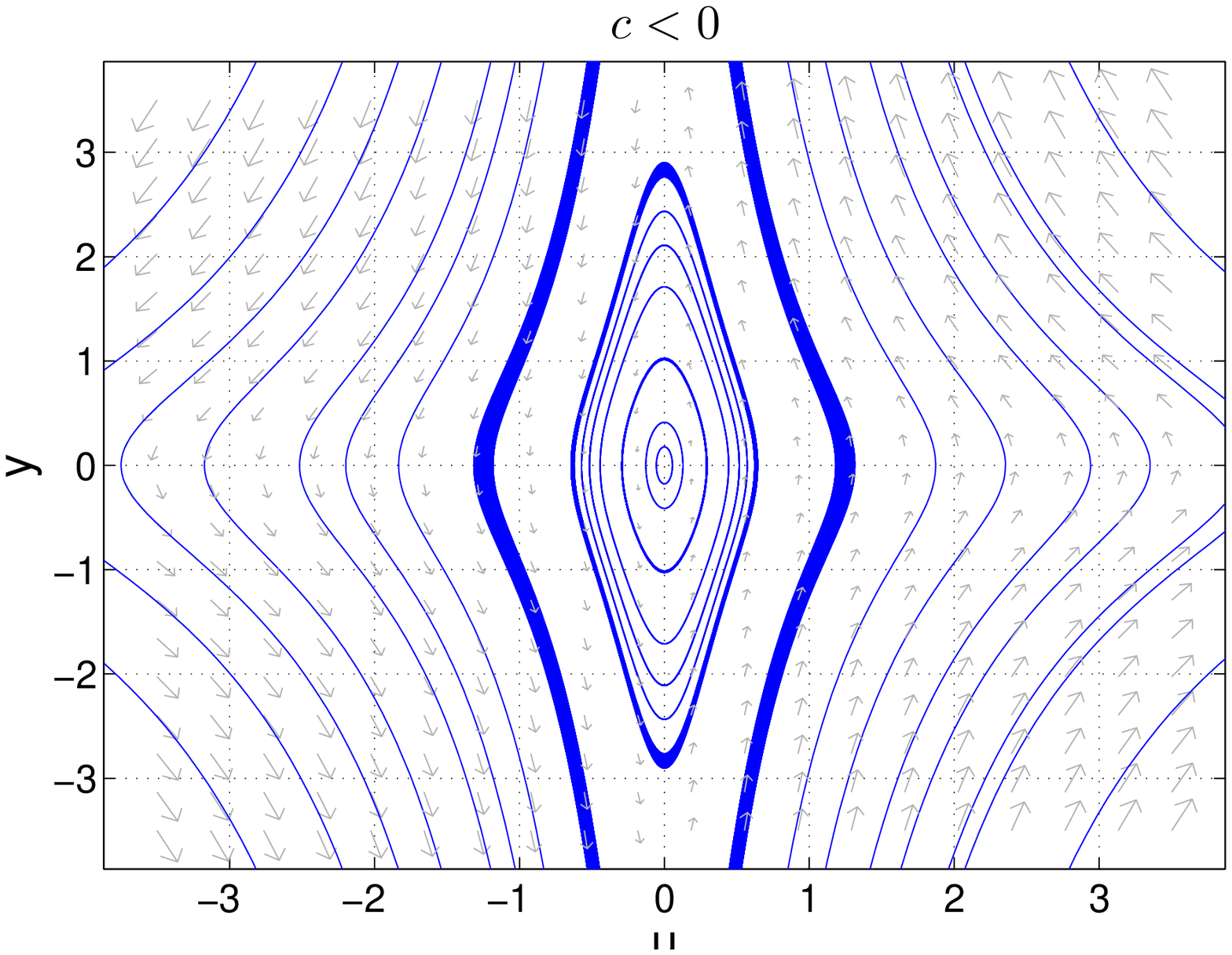}}
\end{center}
\caption{\label{SPE_pp} The phase portraits of system \eqref{SPE_sd_reg}. (a) $c=0.1 > 0$, the singular straight lines $u=u_{s,1}$ and $u=u_{s,2}$ are drawn in green. (b) $c=0$, the singular straight line $u=u_{s,1}\equiv u_{s,2}=0$ is drawn in green. (c) $c=-0.1 < 0$. }
\end{figure}

Let $dz=G(u)d\xi$, the following system:
\begin{equation}\label{SPE_sd_reg}
\left\{\begin{array}{ll}
\displaystyle\frac{du}{d\xi}=yG(u)=y(2c-u^2),\\
\,\\
\displaystyle\frac{dy}{d\xi}=-(G'(u)y^2+F(u))=2u(1+y^2),
\end{array}\right.
\end{equation}
is  the associated regular system of \eqref{SPE_sd}.
The systems of equations in \eqref{SPE_sd} and \eqref{SPE_sd_reg} have the same invariant curve solutions, the main difference between Eqs. \eqref{SPE_sd} and \eqref{SPE_sd_reg} is the parametric representation of the orbit: near $u=u_{s,1}$ and $u=u_{s,2}$, Eq. \eqref{SPE_sd_reg} uses the \textit{fast time variable} $\xi$, while Eq. \eqref{SPE_sd} uses the \textit{slow time variable} $z$ (see \cite{LD07} and \cite{RGC14} for details). Hence, we study the associated regular system of Eq. \eqref{SPE_sd_reg} in order to get the phase portraits of Eq. \eqref{SPE_sd}. Since the first integral of both Eqs. \eqref{SPE_sd} and \eqref{SPE_sd_reg} are the same, thus both of them have the same phase orbits, except on the straight lines $u=u_{s,1}$ and $u=u_{s,2}$. Notice that, for the system \eqref{SPE_sd_reg}, the straight lines $u=u_{s,1}$ and $u=u_{s,2}$ are invariant straight lines.
In Fig.\ref{SPE_pp}(a) and (b), the phase portrait of Eq. \eqref{SPE_sd_reg} are drawn for $c > 0, c=0$ and $c < 0$ respectively.

Let $(u(z), y=u'(z))$ be the parametric representation of an orbit $\gamma$ of the system \eqref{SPE_sd}. Observing the phase portraits when $c>0$ in Fig.\ref{SPE_pp}(a), along the orbit $\gamma$, as $z$ increases
or decreases, the phase point $(u(z), y(z))$ approaches the straight line $u=u_{s,1}=\sqrt{2c}$ (or equivalently $u=u_{s,2}=-\sqrt{2c}$) in the positive direction or the negative direction and $\lim_{u\rightarrow u_{s,1}}|y|=\infty$ (analogously $\lim_{u\rightarrow u_{s,2}}|y|=\infty$). Then, there is a finite value $z=\tilde{z}$ such that:
\begin{equation}\label{SPE_lim}
\lim_{z\rightarrow \tilde{z}}u(z)=u_{s,1} \quad (\rm{or}\ u_{s,2}).
\end{equation}
The profile of the wave defined by $u(z)$ is thus a breaking wave.
In particular, via standard linear stability analysis, we obtain that the point $P\equiv (0, 0)$ is a saddle (the eigenvalues of the linearized system are $\lambda_{1,2}=\pm\sqrt{2c}$),
therefore the stable and unstable manifolds of $P$, which approach the singular straight lines $u=u_{s,1}=\sqrt{2c}$ or  $u=u_{s,2}=-\sqrt{2c}$
give rise to a one sided breaking kink wave solution and a one sided breaking anti-kink wave solution of the SPE (see the appropriate Theorem in \cite{LD07}).
When $c=0$ there is just one straight singular line $u=u_{s,1}\equiv u_{s,2}=0$ containing also the regular equilibrium $P\equiv(0,0)$ and the dynamical behaviour of an orbit is equivalent to that one in the case $c>0$, therefore the solution of the SPE are kink wave solution.
Finally, when $c<0$ the point $P\equiv (0, 0)$ is a center (the eigenvalues of the linearized system are $\lambda_{1,2}=\pm i\sqrt{2c}$) and there are no singular straight lines, therefore a closed loop arise for the traveling wave system \eqref{SPE_sd} for each initial condition.

\section{Regular pulse and front solutions of the SPE: analytic solutions for homoclinic orbits }\label{Sec4}

In this section, we change gears and consider regular pulse and front solutions of the SPE \eqref{SPE_trav}
by calculating convergent, multi-infinite, series solutions for the possible homoclinic orbits of the traveling wave equation \eqref{SPE_trav}.

We employ a recently developed approach \cite{CG13,W09,RGC14}, using the method of undetermined coefficients to derive convergent analytic series for homoclinic orbits of Eq. \eqref{SPE_trav}, corresponding to pulse/front solutions of the SPE \eqref{SPE}.

When $c>0$ the origin is a saddle point of the SPE traveling wave system \eqref{SPE_sd_reg} and a homoclinic orbit arises.
We look for a solution of the following form:
\begin{equation}\label{SPE_hom_orbit}
u(z)=\left\{\begin{array}{lll}
\phi^+(z)\qquad z>0\\
0\qquad\qquad z=0\\
\phi^-(z)\qquad z<0
\end{array}\right.
\end{equation}
where:
\begin{equation}\label{eq4_hom_spsm}
\phi^+(z)=x_0+\sum_{k=\,1}^{\infty} a_k e^{k\alpha z}, \qquad \phi^-(z)=x_0+\sum_{k=\,1}^{\infty} b_k e^{k\beta z},
\end{equation}
and $z=x+ct$, $x_0=0$ is the equilibrium point, $\alpha<0$ and $\beta>0$ are undetermined constants and $a_k, b_k$, with $k\geq 1$, are, at the outset, arbitrary coefficients. Substituting the series \eqref{eq4_hom_spsm} for $\phi^+(z)$
we obtain the following expressions for each term of \eqref{SPE_trav}:
\begin{eqnarray}\label{SPE_terms1}
\phi_{zz}&=&\sum_{k=\,1}^{\infty}a_k(k\alpha)^2e^{k\alpha z},\\\label{SPE_terms2}
\phi^{2}\phi_{zz}&=&\sum_{k=\,3}^{\infty}\sum_{j=\,2}^{k-1}\sum_{l=\,1}^{j-1}a_l a_{j-l}a_{k-j}(k-j)^2\alpha^2e^{k\alpha z},\\\label{SPE_terms3}
\phi\phi^{2}_z&=&\sum_{k=\,3}^{\infty}\sum_{j=\,2}^{k-1}\sum_{l=\,1}^{j-1}a_l a_{j-l}a_{k-j}(j-l)l\alpha^2e^{k\alpha z}.
\end{eqnarray}
Using \eqref{SPE_terms1}-\eqref{SPE_terms3} into the Eq. \eqref{SPE_trav} we obtain:
\begin{equation}\label{SPE_series}
\begin{split}
2\sum_{k=1}^{\infty}(1-c(k\alpha &)^2)a_k e^{k\alpha z}\\
&\,
+\sum_{k=\,3}^{\infty}\sum_{j=\,2}^{k-1}\sum_{l=\,1}^{j-1} ((k-j)^2+2(j-l)l)a_l a_{j-l}a_{k-j}\alpha^2 e^{k\alpha z}=0.
\end{split}
\end{equation}
Comparing the coefficients of $e^{k\alpha z}$ for each $k$, one has for $k=1$:
\begin{equation}\label{eq4_series1}
2(1-c\alpha^2 )a_1=0.
\end{equation}
Assuming $a_1\neq0$ (otherwise $a_k=0$ for all $k>1$ by induction), results in the two possible values of $\alpha$:
\begin{equation}\label{4_series_eig}
\alpha_{1}= \sqrt{\frac{1}{c}},\qquad\qquad \alpha_{2}=-\sqrt{\frac{1}{c}}.
\end{equation}
We are dealing with the case when the equilibrium $x_0=0$ is a saddle, i.e. when $c>0$. In this case, as our series solution \eqref{SPE_hom_orbit} needs to converge for $z >0$, we pick the negative root $\alpha=\alpha_2$.
For $k=2$ we have:
\begin{equation}\label{eq4_series_2}
F(2\alpha_2)a_2=0 \qquad\Rightarrow \qquad a_2=0,
\end{equation}
where $F(k\alpha_2)=2(1-c(k\alpha_2)^2)$.

For $k=3$ we obtain:
\begin{equation}\label{eq4_series_coeff3}
a_3=\frac{3\alpha_2^2a_1^3}{F(3\alpha_2)}.
\end{equation}
For $k>2$ one has:
\begin{equation}\label{eq4_series_coeffk_a}
a_k=\sum_{j=\,2}^{k-1}\sum_{l=\,1}^{j-1} \frac{((k-j)^2+2(j-l)l)a_l a_{j-l}a_{k-j}\alpha_2^2 }{F(k\alpha_2)}.
\end{equation}
Therefore for all $k$ the series coefficients $a_k$ can be iteratively computed in terms of $a_1$:
\begin{equation}\label{eq4_series_coeffk}
a_k=\varphi_k a_1^k,
\end{equation}
where $\varphi_k,\ k>1$ are functions which can be obtained using Eq. \eqref{eq4_series_2}-\eqref{eq4_series_coeffk_a}. They depend on $\alpha_2$ and the constant coefficient $c$ of the Eq. \eqref{SPE_trav}.
Once chosen $\alpha=\alpha_2$ it is easy to see that all the series coefficients $\phi_{2k}$ are equal to zero and the remaining coefficients have the following property:
\begin{equation}\label{prop_coeff}
\phi_{2k+1} {\rm \qquad is\ proportional\ to \qquad }\left(\frac{1}{c}\right)^k,
\end{equation}
therefore, for the convergence of the series coefficients is crucial to control the value of ${c}$.
The first part of the homoclinic orbit corresponding to $z>0$ has thus been determined in terms of $a_1$:         		
\begin{equation}\label{eq4_series_p}
\phi^+ (z)=a_1 e^{\alpha_2 z}+\sum_{k=2}^\infty \varphi_k a_1^k e^{k\alpha_2 z}.
\end{equation}
Notice that the Eq. \eqref{SPE_trav} is reversible under the standard reversibility of classical mechanical systems:
\begin{equation}\label{eq4_norev}
z\rightarrow -z,\qquad   (u, u_x, u_{xx})\rightarrow (u, -u_x, u_{xx}).
\end{equation}
Mathematically, this property would translate to solutions having odd parity in $z$.
Therefore the series solution for $z < 0$ can be easily obtained based on the intrinsic symmetry property of the equation, i.e.:
\begin{equation}\label{eq4_series_m}
\phi^- (z)=-a_1 e^{\alpha_2 z}-\sum_{k=2}^\infty \varphi_k a_1^k e^{k\alpha_2 z}.
\end{equation}
We want to construct a solution continuous at $z=0$, therefore we impose:
\begin{equation}\label{eq4_cont}
a_1+\sum_{k=2}^\infty \varphi_k a_1^k=0.
\end{equation}
Hence we choose $a_1$ as the nontrivial solutions of the above polynomial equation \eqref{eq4_cont}.
In practice the Eq. \eqref{eq4_cont} is numerically solved and the corresponding series solutions are not unique.

\begin{figure}
\begin{center}
\subfigure[] {\epsfxsize=2.5 in\epsfbox{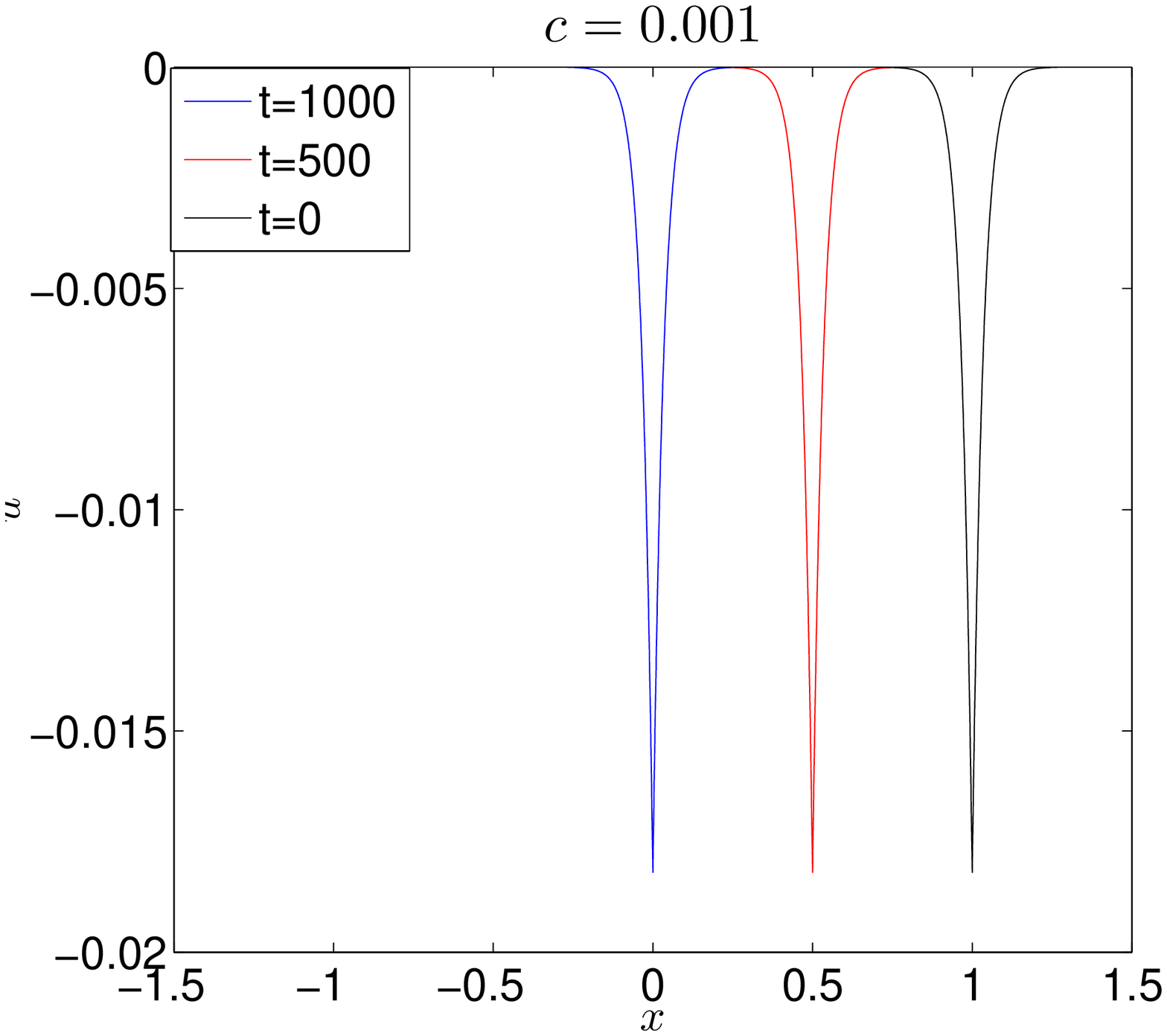}}
\subfigure[] {\epsfxsize=2.65 in \epsfbox{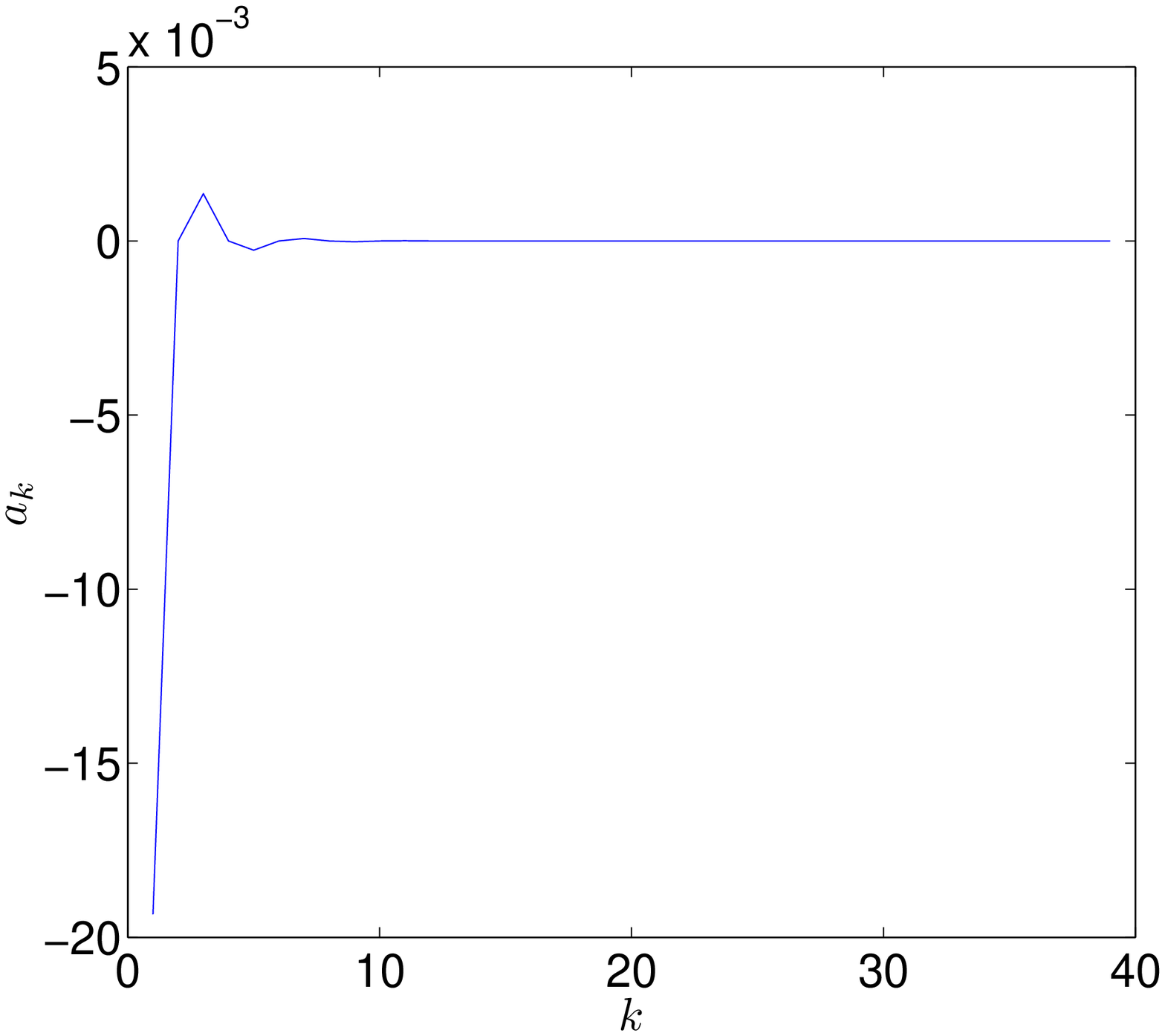}}
\end{center}
\caption{\label{eq4_Fig_hom1} The parameter $c = 0.001$. (a) The series solution $u(z)$ in \eqref{SPE_hom_orbit} for the homoclinic orbit to the saddle point $(0,0)$ plotted as a function of $x$ for different values of $t$, showing traveling wave nature of the solution. Here $a_1 = -0.0193$ is the only solution of the continuity equation \eqref{eq4_cont} truncated at $M=39$. (b) Plot of $a_k$ in \eqref{eq4_series_coeffk_a} versus $k$ shows the series coefficients are converging.}
\end{figure}

Let us now choose $c = 0.001$.
Following the above given computation of the series coefficients, we build the homoclinic orbit to the saddle point $P\equiv(0,0)$. Truncating the series solution up to
$k=39$, the corresponding homoclinic orbit solution is not unique as the continuity condition \eqref{eq4_cont} admits more than one solution. We choose the only value $a_1= -0.0193$ leading to a convergent
series coefficients $a_k$, see Fig.\ref{eq4_Fig_hom1}(b), and the series solution appears as in Fig.\ref{eq4_Fig_hom1}, where also its traveling nature is shown.
We find the continuous solution for the homoclinic orbit shown in Fig.\ref{eq4_Fig_hom1}.
%


\section{The SPE equation with arbitrary coefficients}\label{ASec1}
Let us consider the following short pulse equation with arbitrary coefficients $\beta, \gamma$ :
\begin{equation}
\label{ASPE}
u_{xt}=\beta u+\gamma\frac{1}{6}(u^3)_{xx},
\end{equation}

\noindent where $u = u(x, t)$.
Substituting $u(x,t)=u(x+ct)=u(z)$, where $z=x+ct$ and $c$ is the wave speed, into Eq. \eqref{ASPE} we obtain the following travelling wave equation:
\begin{equation}\label{ASPE_trav}
(\gamma u^2-2c)u_{zz}+2u(\beta+\gamma u_z^2)=0,
\end{equation}
which is equivalent to the following $2$-dimensional traveling wave system:
\begin{equation}\label{ASPE_sd}
\left\{\begin{array}{ll}
\displaystyle\frac{du}{dz}=y,\\
\,\\
\displaystyle\frac{dy}{dz}=\frac{2u(\beta+\gamma y^2)}{2c-\gamma u^2}.
\end{array}\right.
\end{equation}
The system \eqref{ASPE_sd} belongs to the first type of singular traveling wave system \eqref{gen_ld},
where $F$ and $G$ are the following smooth nonlinear functions:
\begin{equation}
\label{ASPE_pos}
F(u)=-2\beta u,\qquad G(u)=2c-\gamma u^2.
\end{equation}
When $c, \gamma >0$ or $c, \gamma <0$ the function $G(u)$ has two real zeros:
\begin{equation}
\label{ASPE_zero_sing}
u_{s,1}=\sqrt{\frac{2c}{\gamma}}\quad {\rm and}\quad u_{s,2}=-\sqrt{\frac{2c}{\gamma}}.
\end{equation}
The function $F(u)$ admits just one zero $u_r=0$ and $F'(u_r)=-2\beta\neq 0$, therefore the point $P\equiv (u_r, 0)\equiv (0, 0)$ is a regular equilibrium of the system \eqref{ASPE_sd}.
Moreover, the second equation in \eqref{ASPE_sd} is discontinuous in the straight lines $u=u_{s,1}=\displaystyle\sqrt{\frac{2c}{\gamma}}$ and $u=u_{s,2}=-\displaystyle\sqrt{\frac{2c}{\gamma}}$.
Being $Y=-\displaystyle\frac{F(u_{s,1})}{G'(u_{s,1})}=-\frac{F(u_{s,2})}{G'(u_{s,2})}=-\displaystyle\frac{\beta}{\gamma}<0$, we have that when $\beta, \gamma>0$ or $\beta, \gamma<0$ there are no singular real equilibria in the singular straight lines $u=u_{s,1}$ and $u=u_{s,2}$, otherwise there exists four critical points $(u_{s,1},\pm \sqrt{Y})$ and $(u_{s,2},\pm {Y})$.
Putting together the above conditions, we obtain that when $c, \gamma >0$ and $\beta<0$ or when $c, \gamma <0$ and $\beta>0$ there exists four critical points $(u_{s,1},\pm \sqrt{Y})$ and $(u_{s,2},\pm {Y})$ on the singular straight lines. In all the other cases there are no real critical points.

Let $dz=G(u)d\xi$, the associated regular system of \eqref{ASPE_sd} is given below:
\begin{equation}\label{ASPE_sd_reg}
\left\{\begin{array}{ll}
\displaystyle\frac{du}{d\xi}=yG(u)=y(2c-\gamma u^2),\\
\,\\
\displaystyle\frac{dy}{d\xi}=-(G'(u)y^2+F(u))=2u(\beta+\gamma y^2).
\end{array}\right.
\end{equation}
We study the associated regular system of Eq. \eqref{ASPE_sd_reg} in order to get the phase portraits of Eq. \eqref{ASPE_sd}, as both of the systems have the same phase orbits, except on the straight lines $u=u_{s,1}$ and $u=u_{s,2}$ (see Section \ref{Sec1} for details).
Via linear stability analysis it is straightforward to obtain that the regular equilibrium $P\equiv (0,0)$ is a saddle when $\beta, c>0$ or $\beta, c>0$, otherwise it is a center. The singular points $(u_{s,1},\pm \sqrt{Y})$ and $(u_{s,2},\pm {Y})$ when exist (i.e. for $c, \gamma >0$ and $\beta<0$ or when $c, \gamma <0$ and $\beta>0$) they are saddle.

\begin{figure}
\begin{center}
\subfigure[] {\epsfxsize=2.3 in \epsfbox{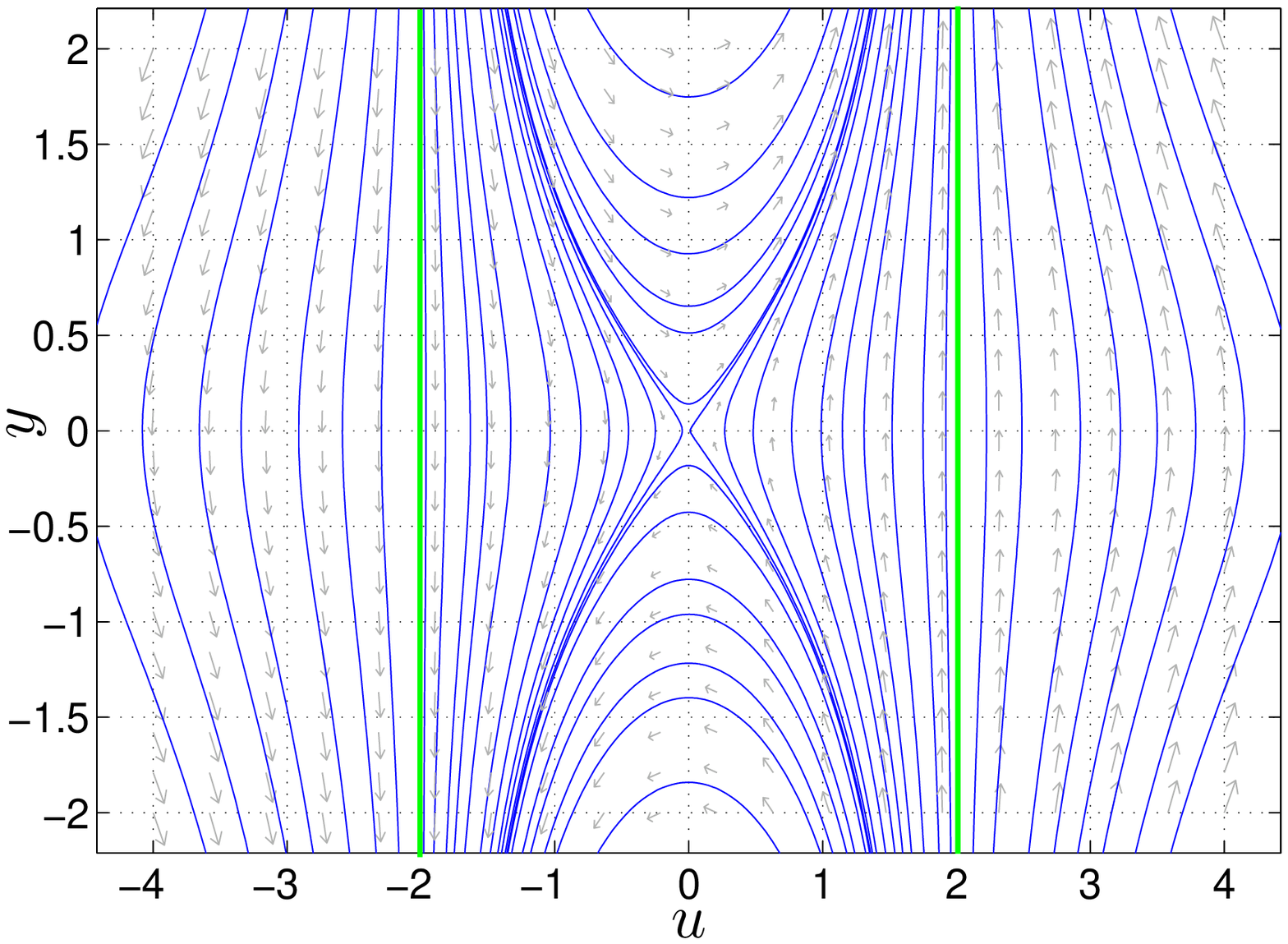}}
\subfigure[] {\epsfxsize=2.3 in \epsfbox{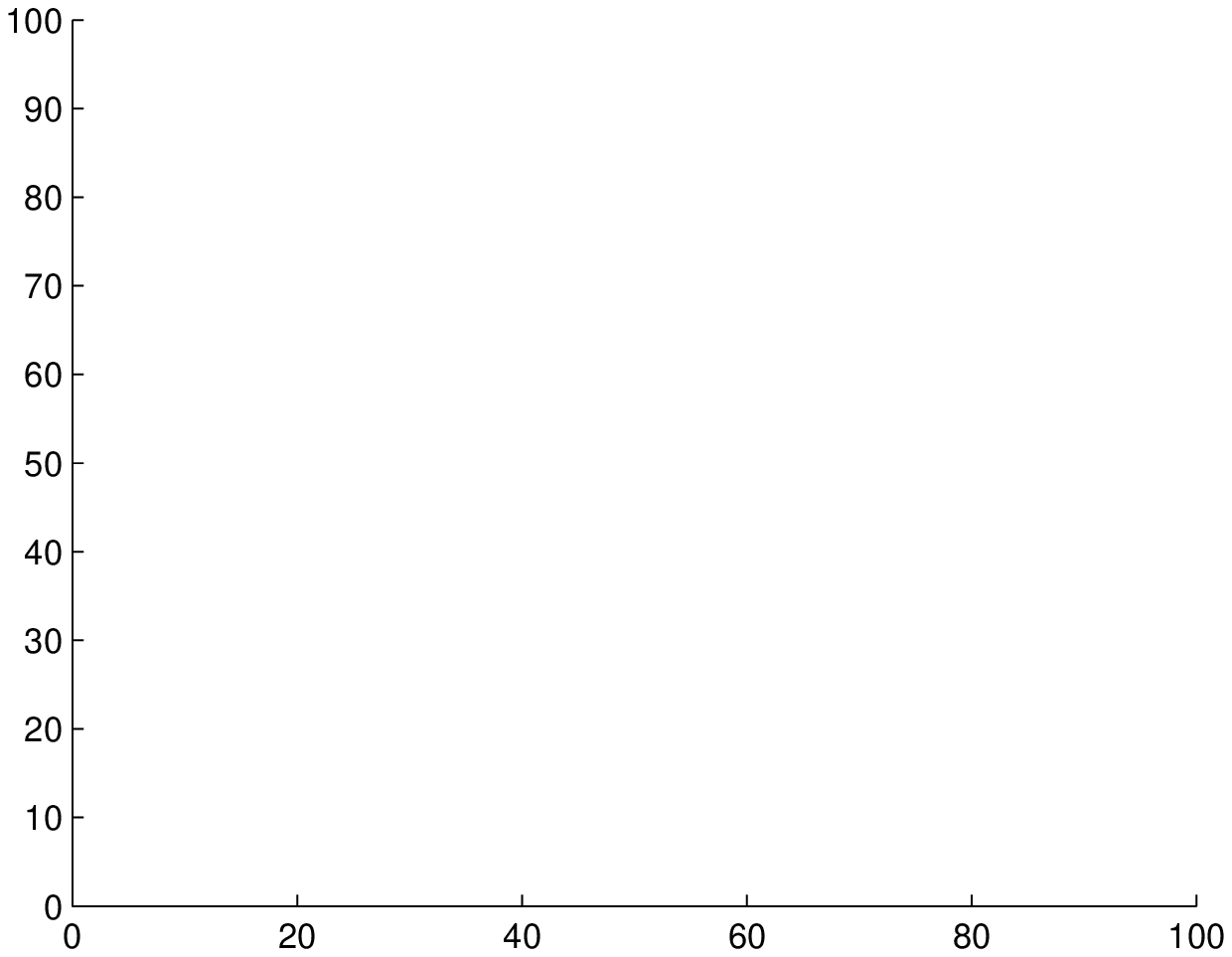}}
\subfigure[] {\epsfxsize=2.3 in \epsfbox{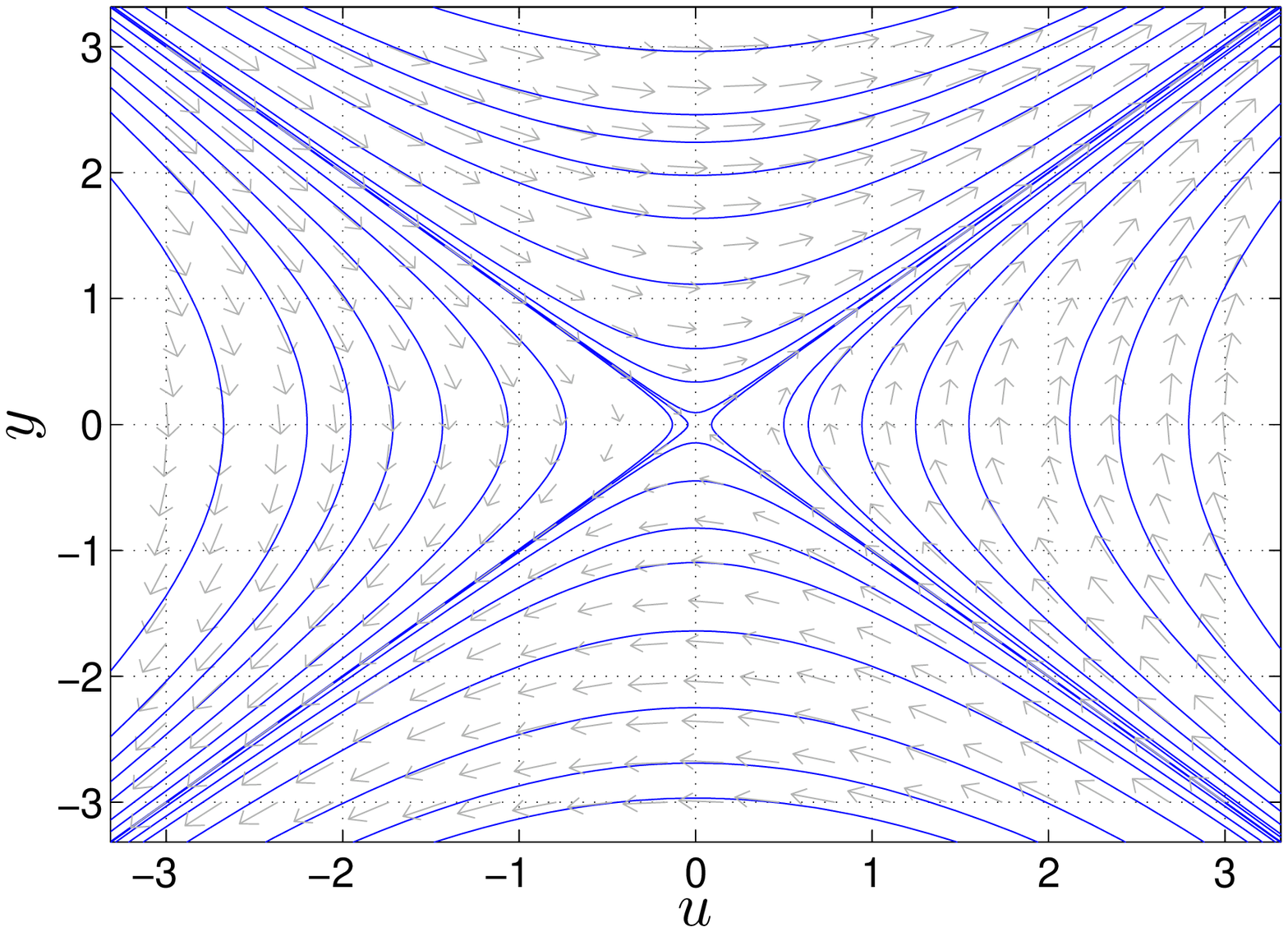}}
\subfigure[] {\epsfxsize=2.3 in \epsfbox{vuoto.eps}}
\subfigure[] {\epsfxsize=2.3 in \epsfbox{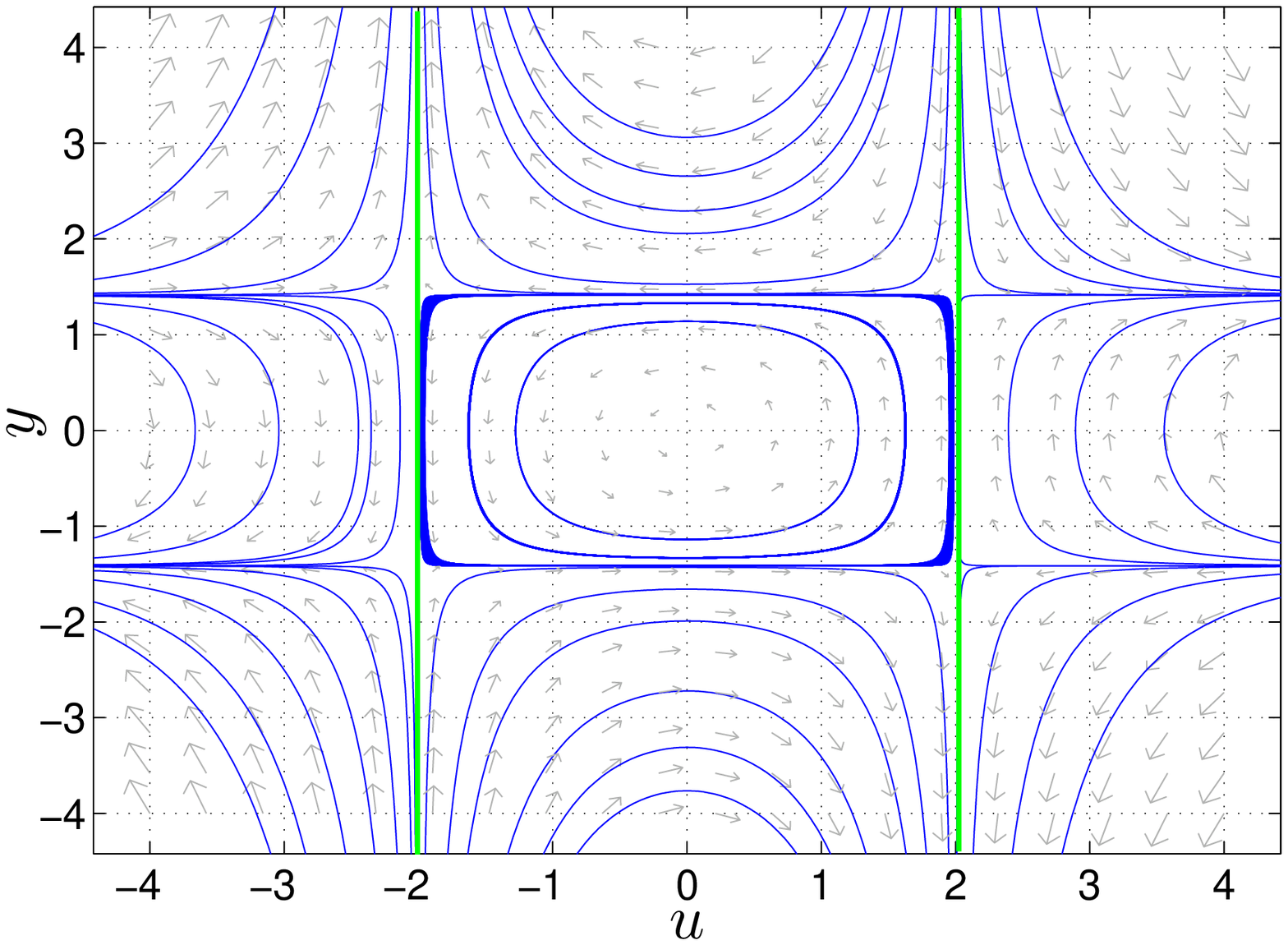}}
\subfigure[] {\epsfxsize=2.3 in \epsfbox{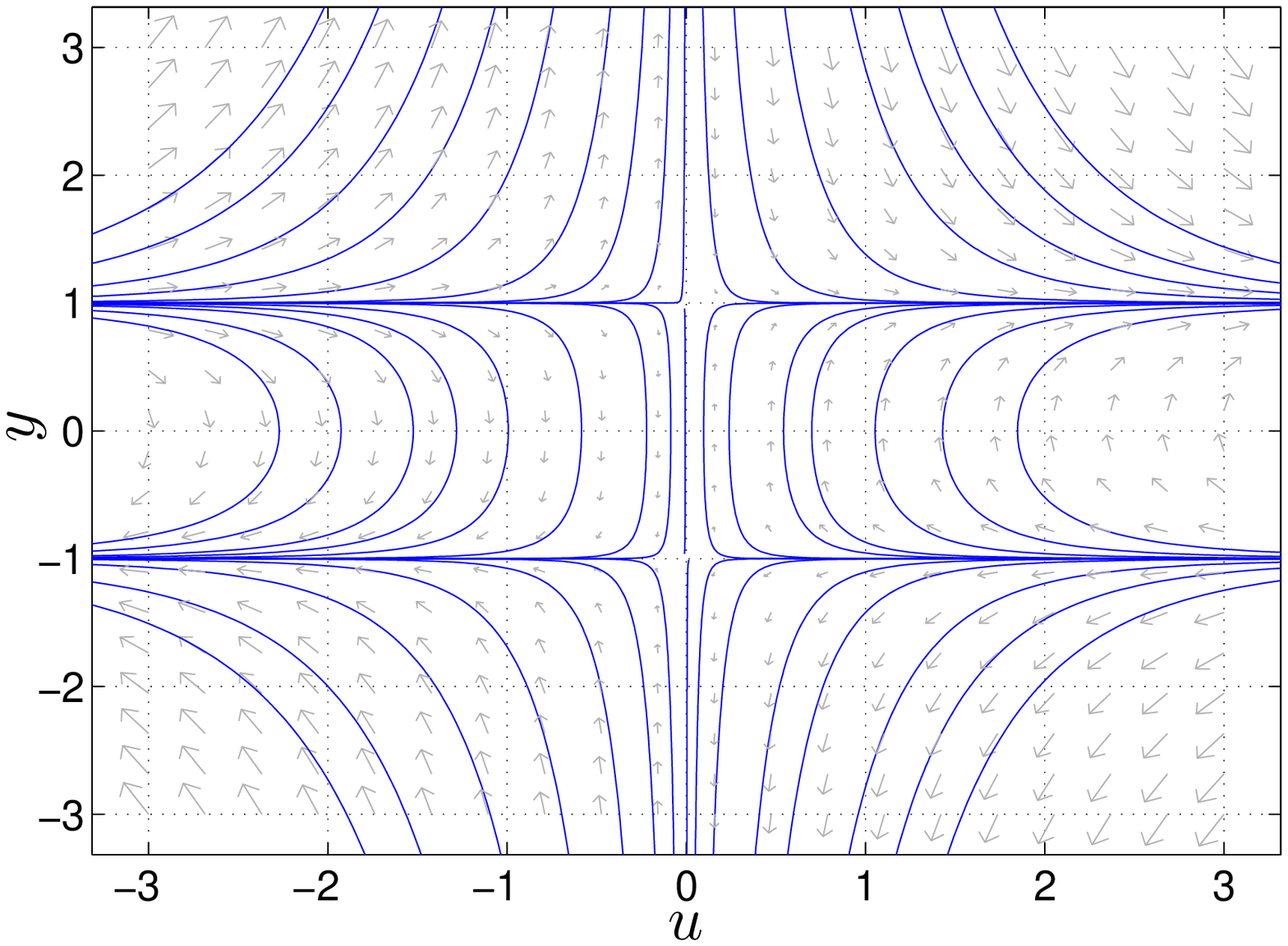}}
\end{center}
\caption{\label{ASPE_pp} The phase portraits of system \eqref{ASPE_sd_reg}. (a) $\beta, c, \gamma > 0$, there are no critical points on the singular straight lines (drawn in green) and the regular equilibrium is a saddle. The phase-portrait is the same as in (a) with $\beta, c, \gamma < 0$.
(b) $\beta, c>0$ and $\gamma < 0$, there are no singular straight lines and the regular equilibrium is a saddle. The phase-portrait is the same as in (b) with $\beta, c<0$ and $\gamma > 0$. (c)  $\beta, c>0$ and $\gamma = 0$, there are no singular straight lines and the regular equilibrium is a saddle. The phase-portrait is the same as in (c) with $\beta, c<0$. (d) $\beta, \gamma>0$ and $c< 0$, there are no singular straight lines and the regular equilibrium is a center. The phase-portrait is the same as in (d) with $\beta, \gamma<0$ and $c> 0$. (e) $\beta>0$ and $c, \gamma <0$, there are four critical points on the two singular straight lines and the regular equilibrium is a center. The phase-portrait is the same as in (e) with $\beta<0$ and $c, \gamma >0$. (f) $\beta>0$, $c<0$ and $\gamma =0$.}
\end{figure}

In Fig.\ref{ASPE_pp}, the phase portrait of Eq. \eqref{ASPE_sd_reg} are drawn for the main dynamical behaviour.

When the system parameters are chosen as in Fig.\ref{ASPE_pp}(a), the profile of the wave defined by $u(z)$ is a breaking wave.
In particular, the stable and unstable manifolds of $P$, being a saddle, approach the singular straight lines $u=u_{s,1}=\sqrt{2c}$ or  $u=u_{s,2}=-\sqrt{2c}$
and give rise to a one sided breaking kink wave solution and a one sided breaking anti-kink wave solution of the SPE with arbitrary coefficients.
When the parameters are as in Fig.\ref{ASPE_pp}(d), the point $P\equiv (0, 0)$ is a center and there are no singular straight lines, therefore a closed loop arise for the traveling wave system \eqref{ASPE_sd} for each initial condition.

Finally, for the parameters as in Fig.\ref{ASPE_pp}(e) periodic cusp wave solutions (both of peak and valley type) arise (see also \cite{ZCH06}).

\section{Regular pulse and front solutions of the SPE with arbitrary coefficients: analytic solutions for homoclinic orbits }\label{ASec4}

In this section, we use the same approach as in Section \ref{Sec4} to compute convergent, multi-infinite series solutions for the possible homoclinic orbits of the traveling wave equation \eqref{ASPE_trav}.

When $\beta, c>0$ or $\beta, c>0$ the origin is a saddle point of the SPE traveling wave system \eqref{ASPE_sd_reg} and a homoclinic orbit arises.For this choice of the system parameters, we look for a solution of
the Eq. \eqref{ASPE_trav} with the same form as in \eqref{SPE_hom_orbit}.
Substituting the series \eqref{eq4_hom_spsm} for $\phi^+(z)$ in \eqref{ASPE_trav},
we obtain the following equation:
%
\begin{equation}\label{ASPE_series}
\begin{split}
2\sum_{k=1}^{\infty}(\beta-c(k&\alpha)^2)a_k e^{k\alpha z}
\\
&\,
+\gamma\sum_{k=\,3}^{\infty}\sum_{j=\,2}^{k-1}\sum_{l=\,1}^{j-1} ((k-j)^2+2(j-l)l)a_l a_{j-l}a_{k-j}\alpha^2 e^{k\alpha z}=0,
\end{split}
\end{equation}
and comparing the coefficients of $e^{k\alpha z}$ for each $k$, we obtain for $k=1$:
\begin{equation}\label{Aeq4_series1}
2(\beta-c\alpha^2 )a_1=0.
\end{equation}
Assuming $a_1\neq0$ (otherwise $a_k=0$ for all $k>1$ by induction), results in the two possible values of $\alpha$:
\begin{equation}\label{A4_series_eig}
\alpha_{1}= \sqrt{\frac{\beta}{c}},\qquad\qquad \alpha_{2}=-\sqrt{\frac{\beta}{c}}.
\end{equation}
As we are dealing with the case when the equilibrium $x_0=0$ is a saddle (i.e. $\beta, c>0$ or $\beta, c>0$), the values in \eqref{A4_series_eig} are real and opposite. The series solution \eqref{SPE_hom_orbit} has to converge, therefore we pick the negative root $\alpha=\alpha_2$ for $z >0$.

For $k=2$ we obtain:
\begin{equation}\label{Aeq4_series_2}
F(2\alpha_2)a_2=0 \qquad\Rightarrow \qquad a_2=0,
\end{equation}
where $F(k\alpha_2)=2(\beta-c(k\alpha_2)^2)$.
For $k>2$ one has:
\begin{equation}\label{Aeq4_series_coeffk_a}
a_k=\gamma\sum_{j=\,2}^{k-1}\sum_{l=\,1}^{j-1} \frac{((k-j)^2+2(j-l)l)a_l a_{j-l}a_{k-j}\alpha_2^2 }{F(k\alpha_2)},
\end{equation}
therefore, using Eq. \eqref{Aeq4_series_2}-\eqref{Aeq4_series_coeffk_a}, the series coefficients $a_k, \forall k>1$ can be iteratively computed in terms of $a_1$, and they can be written in the same form given in \eqref{eq4_series_coeffk}.
Once chosen $\alpha=\alpha_2$ it is easy to see that all the series coefficients $\phi_{2k}$ are equal to zero and the remaining coefficients have the following property:
\begin{equation}\label{Aprop_coeff}
\phi_{2k+1} {\rm \qquad is\ proportional\ to \qquad }\left(\frac{\gamma}{c}\right)^k,
\end{equation}
therefore, for the convergence of the series coefficients is crucial to control the quotient $\displaystyle\frac{\gamma}{c}$.
The first part of the homoclinic orbit corresponding to $z>0$ has thus been determined as in \eqref{eq4_series_p}, where the coefficients $a_k$ are given in \eqref{Aeq4_series_coeffk_a} and
the second part for $z<0$ is rapidly obtained as in \eqref{eq4_series_m} thanks to the reversibility property of the Eq.\eqref{ASPE_trav}.       		 
%
Finally, the value of $a_1$ can be computed as the solution of the continuity equation \eqref{eq4_cont}. 

\begin{figure}
\begin{center}
\subfigure[] {\epsfxsize=2.6 in\epsfbox{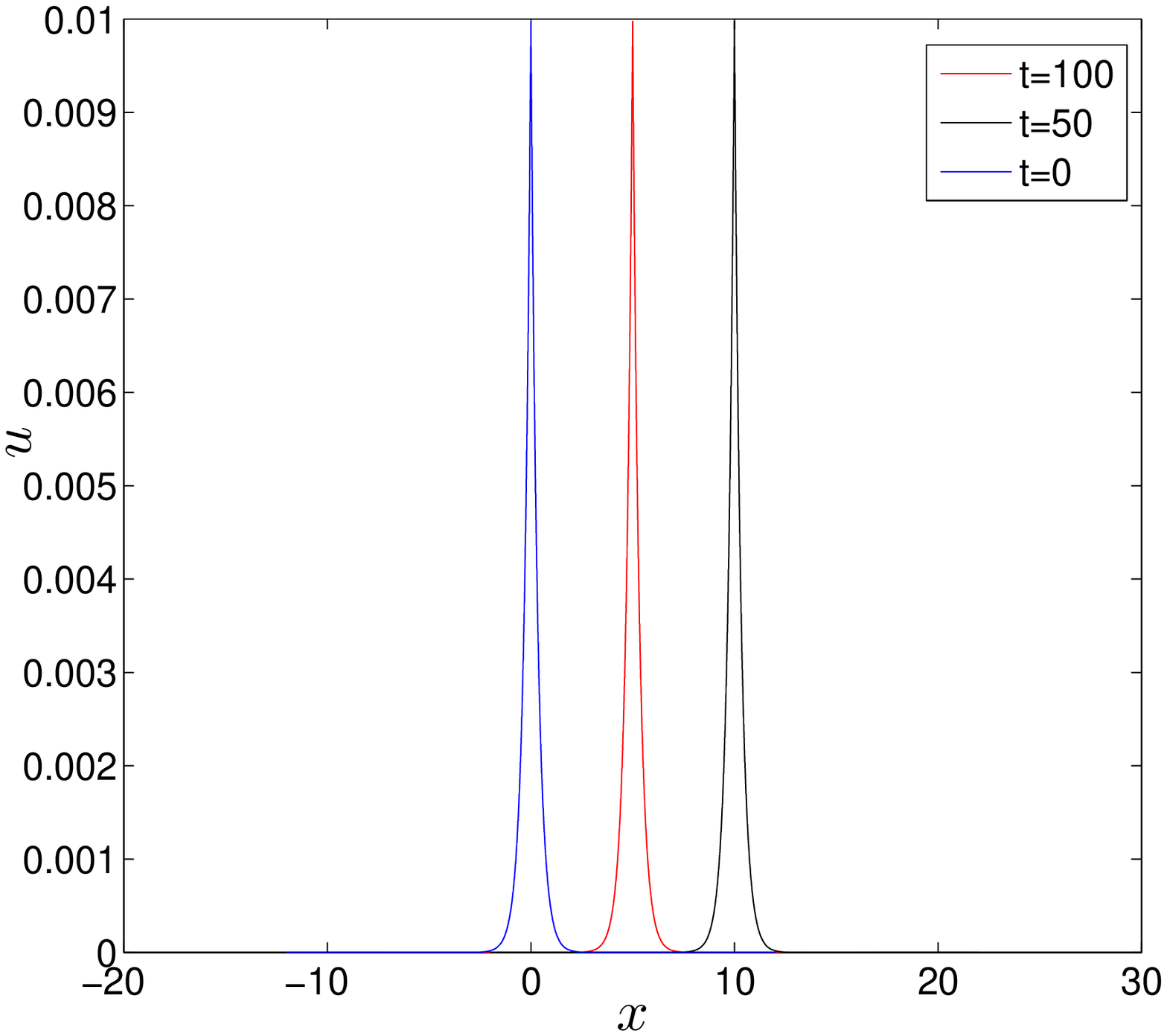}}
\subfigure[] {\epsfxsize=2.6 in \epsfbox{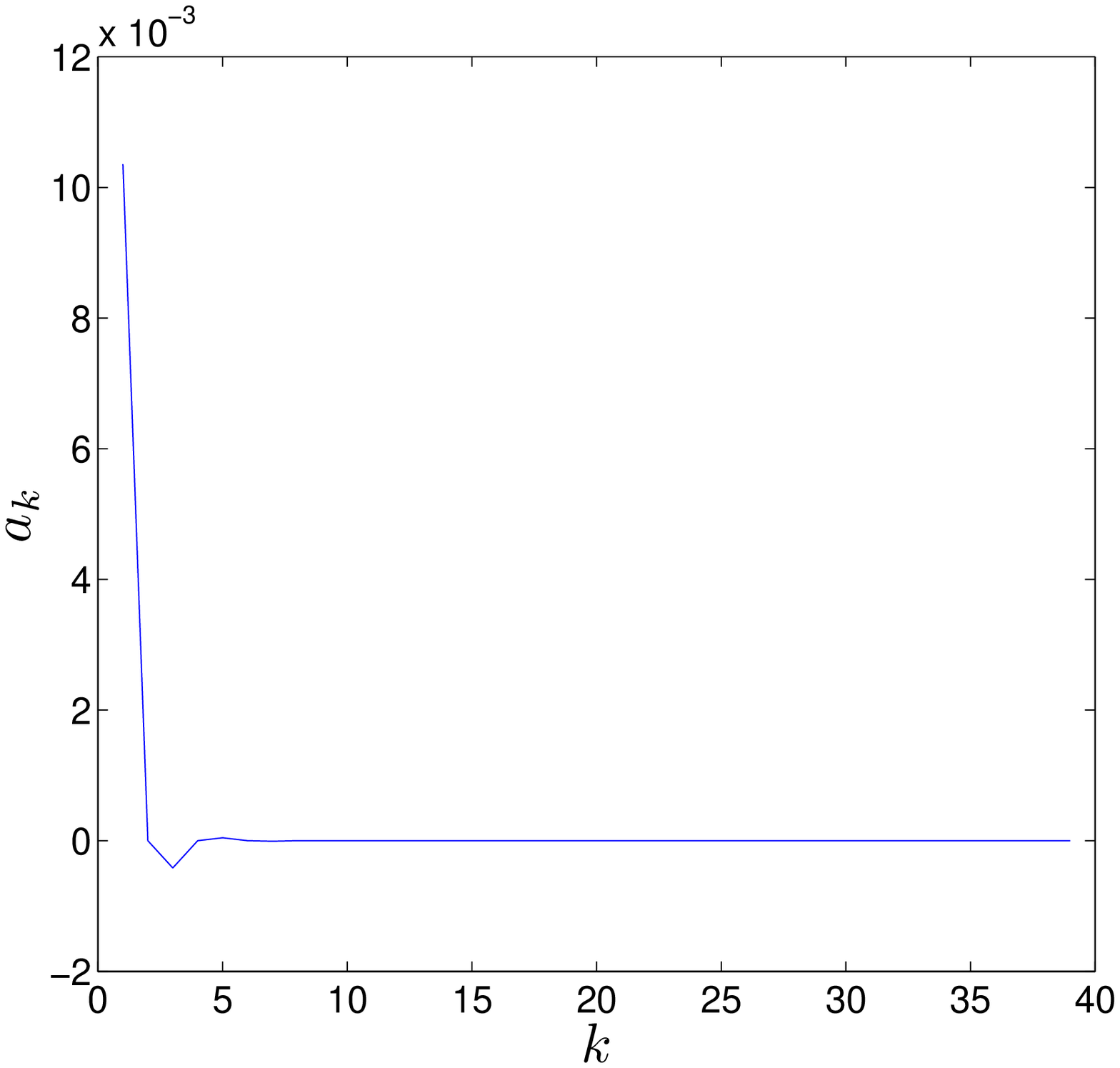}}
\end{center}
\caption{\label{Aeq4_Fig_hom1} The parameters $c = 0.1, \beta=1$ and $\gamma=200$. (a) The series solution $u(z)$ in \eqref{SPE_hom_orbit} for the homoclinic orbit to the saddle point $(0,0)$ of the Eq.\eqref{ASPE_trav} plotted as a function of $x$ for different values of $t$, showing traveling wave nature of the solution. Here $a_1= 0.0104$ is the only solution of the continuity equation \eqref{eq4_cont} truncated at $M=39$. (b) Plot of $a_k$ in \eqref{Aeq4_series_coeffk_a} versus $k$ shows the series coefficients are converging.}
\end{figure}

\begin{figure}
\begin{center}
\subfigure[] {\epsfxsize=2.6 in \epsfbox{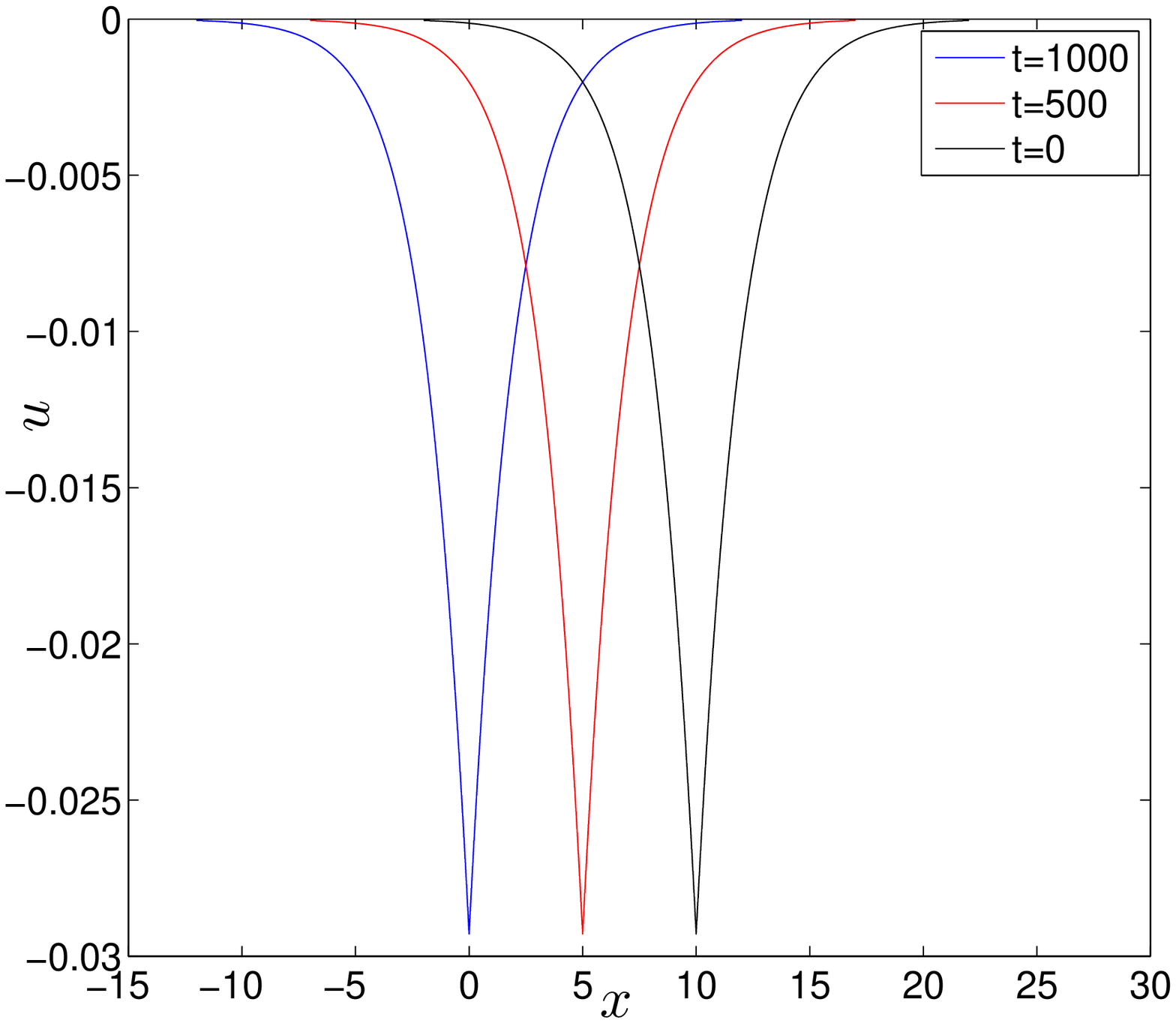}}
\subfigure[] {\epsfxsize=2.6 in \epsfbox{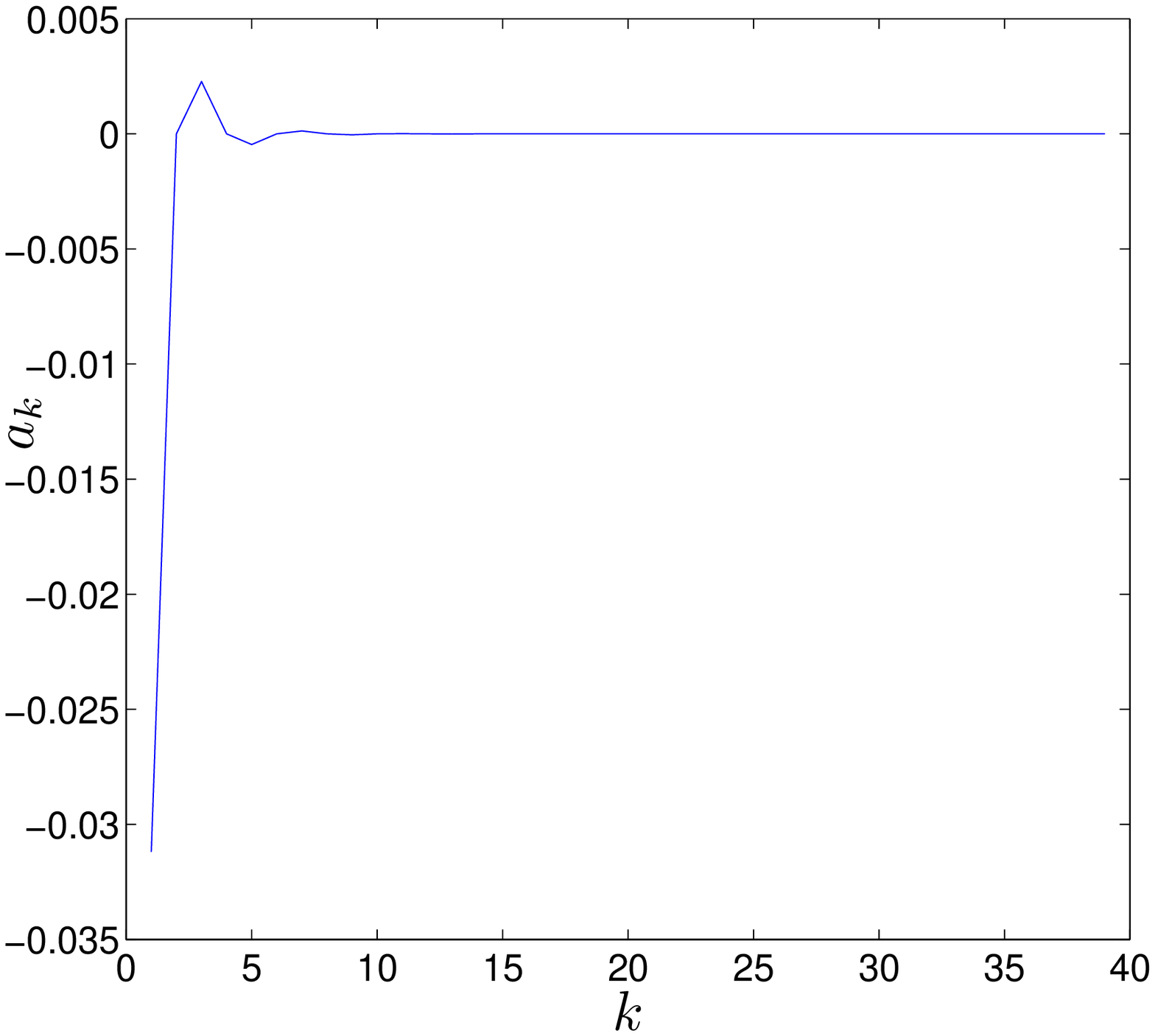}}
\end{center}
\caption{\label{Aeq4_Fig_hom2} The parameters $c = 0.01, \beta=0.003$ and $\gamma=4$. The continuity equation \eqref{eq4_cont} truncated at $M=39$ does not admit a unique solution. Here we choose the solution $a_1= -0.0312$. (a) The series solution $u(z)$ in \eqref{SPE_hom_orbit} for the homoclinic orbit to the saddle point $(0,0)$ of the Eq.\eqref{ASPE_trav} plotted as a function of $x$ for different values of $t=0$. (b) Plot of $a_k$ in \eqref{Aeq4_series_coeffk_a} versus $k$ shows the series coefficients converge.}
\end{figure}

Let us choose $c = 0.1, \beta=1$ and $\gamma=200$.
Following the above given computation of the series coefficients, we build the homoclinic orbit to the saddle point $P\equiv(0,0)$. Truncating the series solution up to
$k=39$, the continuity condition \eqref{eq4_cont} admits more than one solution; we choose the only value $a_1= 0.0104$ leading to a convergent
series coefficients $a_k$, see Fig.\ref{Aeq4_Fig_hom1}(b), and the series solution appears as in Fig.\ref{Aeq4_Fig_hom1}, where also its traveling nature is shown.
In the second numerical example, we choose $c = 0.01, \beta=0.003$ and $\gamma=4$. Again the solution is not unique as the continuity condition admits more than one solution. We choose $a_1= -0.0312$  to obtain both the convergence of the series coefficients and the continuity at the origin, as shown in Fig.\ref{Aeq4_Fig_hom2}.

\section{Lagrangian via Jacobi's Last Multiplier}\label{Sec2}

In this section, we derive a Lagrangian for the traveling wave equation \eqref{SPE_trav} of the SPE equation.
While this may be done by simply matching the terms in this equation to those in the Euler-Lagrange equation,
we use an alternative approach here using the technique of Jacobi's Last Multiplier. In the next section,
this Lagrangian will be employed to construct solitary wave solutions of the SPE equation, having amplitude
and width parameters optimized to satisfy the corresponding Euler-Lagrange equations.

Jacobi \cite{Jacobi} first described his method for the ``Last Multiplier''
(which we shall refer to as the Jacobi Last Multiplier, or JLM for short) in Konigsberg over $1842-1843$.
It essentially yields an extra first integral for dynamical systems by locally reducing an $n$-dimensional
system to a two-dimensional vector field on the intersection of the $n-2$ level sets formed by the first integrals.
After the work of Jacobi, the JLM received a fair amount of attention, including in a classic paper by Sophus Lie \cite{Lie}
placing it within his general framework of infinitesimal transformations. In 1874 Lie \cite{Lie} showed that one could use point
symmetries to determine last multipliers. A clear formulation in terms of solutions or first
integrals and symmetries is given by L.Bianchi \cite{Bianchi}.

Subsequently it was used for computing first integrals of some ordinary differential equations (ODEs).
The relation between the Jacobi multiplier denoted by $M$, and the Lagrangian $L$ for any second-order
ODE was derived by Rao \cite{Rao}, following some investigations in the early twentieth century \cite{Wit88}.
After Rao's work, the JLM does not appear to have been  extensively employed in work on dynamical systems
till it was recently used by Leach and Nucci to derive Lagrangians for a variety of ODE systems \cite{NL05,NL08,NT10}.
Recently more geometric formulation of JLM has been studied in \cite{CGK09}.

\vskip 1.5cm
In this section, we first use the JLM to derive  a Lagrangian for the traveling-wave equation of the SPE equation. And then
we also investigate possible isochronous behavior in this traveling-wave equation (corresponding to singly-periodic
wavetrains of the SPE PDE), we therefore also attempt to map the potential term to either the simple harmonic oscillator (SHO) or the isotonic potential for specific values of the coefficient parameters of the SPE equation.

\subsection{Derivation of the Lagrangian via the JLM}\label{sec0a}
Given a $m$-dimensional system of first order ODEs $y_i^{'} = f_i(x, y_i), i=1,\dots, m$, the Jacobi last multiplier, denoted by $M(x, y_i)$, is defined as
an integrating factor of the system satisfing the following equation:
\begin{equation}\label{1ODEM}
\frac{d(\log M)}{dx}+ \sum_{i=1}^m\frac{\partial f_i(x, y_i)}{\partial y_i} = 0
\end{equation}
Since a second-order ODE $y^{''} = f(x, y, y^{'})$ is equivalent to a $2$-dimensional system of first order ODEs, the corresponding Jacobi multiplier $M(x, y, y^{'})$ satisfies the following equation:
\begin{equation}\label{2ODEM}
\frac{d(\log M)}{dx}+ \frac{\partial f(x, y,y^{'})}{\partial y^{'}} = 0,
\end{equation}
see for details \cite{NL02,NL04,NT10,N07,TGCsub}.

Let us rewrite the Euler-Lagrange equation:
\begin{equation}\label{EL}
\frac{d}{dx}\left(\frac{\partial L}{\partial y^{'}}\right) = \frac{\partial L}{\partial y},
\end{equation}
by inserting $y^{''} = f(x, y, y^{'})$ as follows:
\begin{equation}\label{ELn}
\frac{\partial^2 L}{\partial x\partial y^{'}} +y^{'}\frac{\partial^2 L}{\partial y\partial y^{'}}+f(x, y, y^{'}) \frac{\partial^2 L}{\partial y^{'2}}= \frac{\partial L}{\partial y}.
\end{equation}
Assuming $\displaystyle\frac{\partial^2 L}{\partial y^{'2}}\neq 0$ and differentiating equation \eqref{ELn} with respect to $y^{'}$, the following equation is obtained:
\begin{equation}\label{ELnn}
\frac{d}{dx}\log\left(\frac{\partial^2 L}{\partial y^{'2}}\right)+\frac{\partial f}{\partial y^{'}}=0.
\end{equation}
Comparing the equation \eqref{ELnn} with \eqref{2ODEM}, we find the equation which connects the JLM to the Lagrangian $L$ \cite{Rao,Wit88,NL02}:
\begin{equation}\label{cML}
M=\displaystyle\frac{\partial^2 L}{\partial y^{'2}}.
\end{equation}
Therefore, the appropriate Lagrangian for the system can be determined starting from the JLM.

\subsection{Search for isochronous behavior via the JLM}\label{sec0b}
Isochronous systems, whose motions are periodic with a single period in extended regions of phase-space (often the entire phase-space)
have attracted significant interest in recent years, especially following the work of Calogero and his collaborators
(see \cite{Cal11,Cal12} and references therein), which revealed the near-ubiquity of such dynamics ``close'' to
numerous classes of dynamical systems.
In addition, in \cite{CV05} it is proved that, up to a possible translation and the addition of a constant,
planar polynomial systems exhibiting
isochronicity are described by either the linear SHO potential $V(x) = \displaystyle\frac{ \omega^2 x^2 }{2}$,
or the isotonic potential
$V(x) = \displaystyle\frac{ \omega^2 x^2 }{8} +\frac{ c^2  }{x^2}$. These are rational potential functions,
and systems which may be mapped to them exhibit oscillatory solutions with the same period
$T =\displaystyle\frac{2\pi}{\omega}$. Irrational potentials,
such as some with discontinuous second derivatives, may also be isochronous.

In \cite{C07,HLP07} Chouikha and Hill et al. studied conditions under which the so-called \textit{Cherkas} system \cite{C76} with a
center at the origin as well as a five-parameter of reversible cubic systems may exhibit isochronicity. However, the study of the
isochronicity conditions
is non-trivial, and the technique required considerable computational effort. The same problem was re-examined in \cite{CG10,GC11} using the JLM
to derive the conditions for isochronous solution behavior much more directly and with far less computational effort (see \cite{GC13}
for complete review). Here we shall follow this latter approach to examine \eqref{SPE_trav} for possible isochronous behavior.

Once derived a Lagrangian via the use of the JLM, the next step is to attempt a transformation of variables which might map the
Hamiltonian to that of the linear SHO or the isotonic potential. As discussed above, such a mapping would prove isochronous
behavior of the original dynamical system
\cite{CV05}.

\subsection{Lagrangian for the SPE traveling-wave equation}\label{sec2}

In this subsection, we consider \eqref{SPE_trav} for the traveling waves of the SPE equation.
To compute the JLM, we use the equation \eqref{2ODEM} which, for the system \eqref{SPE_trav}, becomes:
\begin{equation}\label{anhaM}
\frac{d(\log M)}{dz} - \frac{ 4 u}{u^2 -2 c }u^{'} = 0.
\end{equation}
The solution of the equation \eqref{anhaM} is given by:

\begin{equation}\label{anhaMsol}
M(u)=(u^2 - 2c)^2.
\end{equation}
Using the equation \eqref{cML}, we find the appropriate Lagrangian for \eqref{SPE_trav} to be:
\begin{equation}\label{anhaL}
L(u, u^{'})=\frac{1}{2}M(u)u^{'2}-V(u),
\end{equation}
where the potential energy $V(u)$ satisfies the following equation:
\begin{equation}\label{anhaV}
V^{'}(u)= 2 u (u^2 - 2 c).
\end{equation}
Integrating, we obtain $V(u) = \displaystyle\frac{u^4}{2} - 2 c u^2$.

Applying a Legendre transformation to the Lagrangian $L$ in \eqref{anhaL}, one can find the corresponding Hamiltonian to be:
\begin{equation}\label{anhaH}
H=\frac{1}{2}\left(\frac{p}{\sqrt{M}}\right)^2+V(u),
\end{equation}
where the conjugate momentum $p=\displaystyle\frac{\partial L}{\partial u^{'}}=M(u)u^{'}$.

Next, let us search for isochronous behavior via the use of the JLM. If such behavior were found, it would correspond to period traveling
wavetrains in the SPE equation.

Define the canonical variables:
\begin{equation}\label{can_var}
P=\frac{p}{\sqrt{M}}\quad {\rm and}\quad Q=Q(u)
\end{equation}
to be some function of $u$ such that the Poisson bracket $[P, Q]=[p, q]$ is invariant. This implies that $Q^{'}(u)=\sqrt{M(u)}$. Assuming that there exists a linearizing transformation such that $V(u)\rightarrow Q(u)^2/2$, implies that $V'(u)=Q(u)Q'(u)=\sqrt{M(u)}Q(u)$, so that:
\begin{equation}\label{Q1}
Q(u)=\frac{V'(u)}{\sqrt{M(u)}}=2u.
\end{equation}
Integrating $Q^{'}(u)=\sqrt{M(u)}$, we obtain the following equation:
\begin{equation}\label{QQ1}
Q(u) = \frac{u^3}{3} - 2 c u.
\end{equation}
Since we cannot obtain the same value for $Q$ from Eqs.\eqref{Q1} and \eqref{QQ1}, it cannot be a canonical variable. Thus, the potential cannot be directly mapped to a linear harmonic oscillator. Thus, at least
within the framework of this method, we do not find any parameter sets for the paramter $c$ for which the SPE traveling-wave equation has isochronous
solutions corresponding to singly-periodic traveling wavetrains of the SPE NLPDE.

Hence, we turn next to the construction of solitary waves of the SPE equation using a variational approach.

\section{Variational Formulation}

\subsection{The variational approximation for regular solitons}

The procedure for constructing regular solitary waves with exponentially decaying tails is well-known. It is widely employed in many areas of Applied Mathematics and goes by the name of the Rayleigh-Ritz method. In this section, we shall employ it to construct regular solitary waves of \eqref{SPE_trav}.

The localized regular solitary wave solutions will be found by assuming a Gaussian trial function \eqref{gtf}, and substitutiting this into the Lagrangian \eqref{anhaL}:
\begin{equation}\label{gtf}\phi=A\exp\left(-\frac{z^2}{\rho^2}\right).\end{equation}
Note that it is standard to use such Gaussian ansatz\"en for analytic tractability. This is true even for simpler nonlinear PDEs where exact solutions may be known, and have the usual $sech$ or $sech^2$ functional forms. The exponential trial function typically captures these more exact solitary wave forms extremely well in the core or central region of the soliton, with the two often being indistinguishable
when plotted together. However, the accuracy is typically somewhat worse in the tails, sometimes with errors of upto a few percent there.

\begin{figure}
\begin{center}
{\epsfxsize=3 in \epsfbox{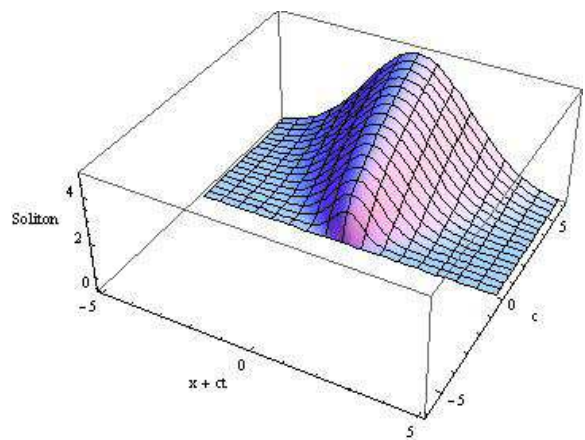}}
\end{center}
\caption{\label{fig_soliton} The regular soliton plotted for different values of $c$.}
\end{figure}

Next, substituting the trial function into the Lagrangian and integrating over all space yields the following ``averaged Lagrangian'' or action:
\begin{equation}\label{action1}\frac{A^2\sqrt{\pi}}{36\rho}(-9A^2\rho^2+36\sqrt{2} (c^2 + c \rho^2) + \sqrt{6}A^4 - 18 c A^2).\end{equation}
The next step is to optimize the trial functions by varying the action with respect to the trial function parameters, viz. the core amplitude $A$, and the core width $\rho$. This determines the optimal parameters for the trial function or solitary wave solution, but within the particular functional form chosen for the trial function ansatz, in this case a Gaussian. The resulting variational Euler-Lagrange equations, by varying $A$ and $\rho$ respectively, are the system of algebraic equations:
\begin{eqnarray}\label{var_eq}\rho^2(3\sqrt{2}a_1+A(2\sqrt{3}a_2+3Aa_3))&=&3\sqrt{2}(1+\rho^2),\\
\rho^2(18\sqrt{2}a_1+A(8\sqrt{3}a_2+9Aa_3))&=&18\sqrt{2}(-1+\rho^2).\end{eqnarray}
Given their relative simplicity, and assuming $a_1=1/2$, $a_3=1$, a nontrivial solution to the equations \eqref{var_eq} is the following:
\begin{eqnarray}\label{solA}A&=&\frac{4(10-\rho^2)a_2}{3\sqrt{3}(\rho^2-6)},\\
\label{solrho}\rho^2&=&\frac{80a_2^2+2\sqrt{2}(81-4\sqrt{81\sqrt{2}a_2^2+50a_2^4})}{27\sqrt{2}+16a_2^2}.\end{eqnarray}
The optimized variational soliton for the regular solitary waves of the traveling-wave equation \eqref{SPE_trav} is given by the trial function \eqref{gtf} with $A$ and $\rho$ respectively given in \eqref{solA} and \eqref{solrho}. Figure \ref{fig_soliton} shows the resulting regular solitary wave solution for various values of the parameter $c$. Note that the tail analysis revealed the need for $a_1<1$ in regimes with regular solitary waves.

\begin{figure}
\begin{center}
{\epsfxsize=3 in \epsfbox{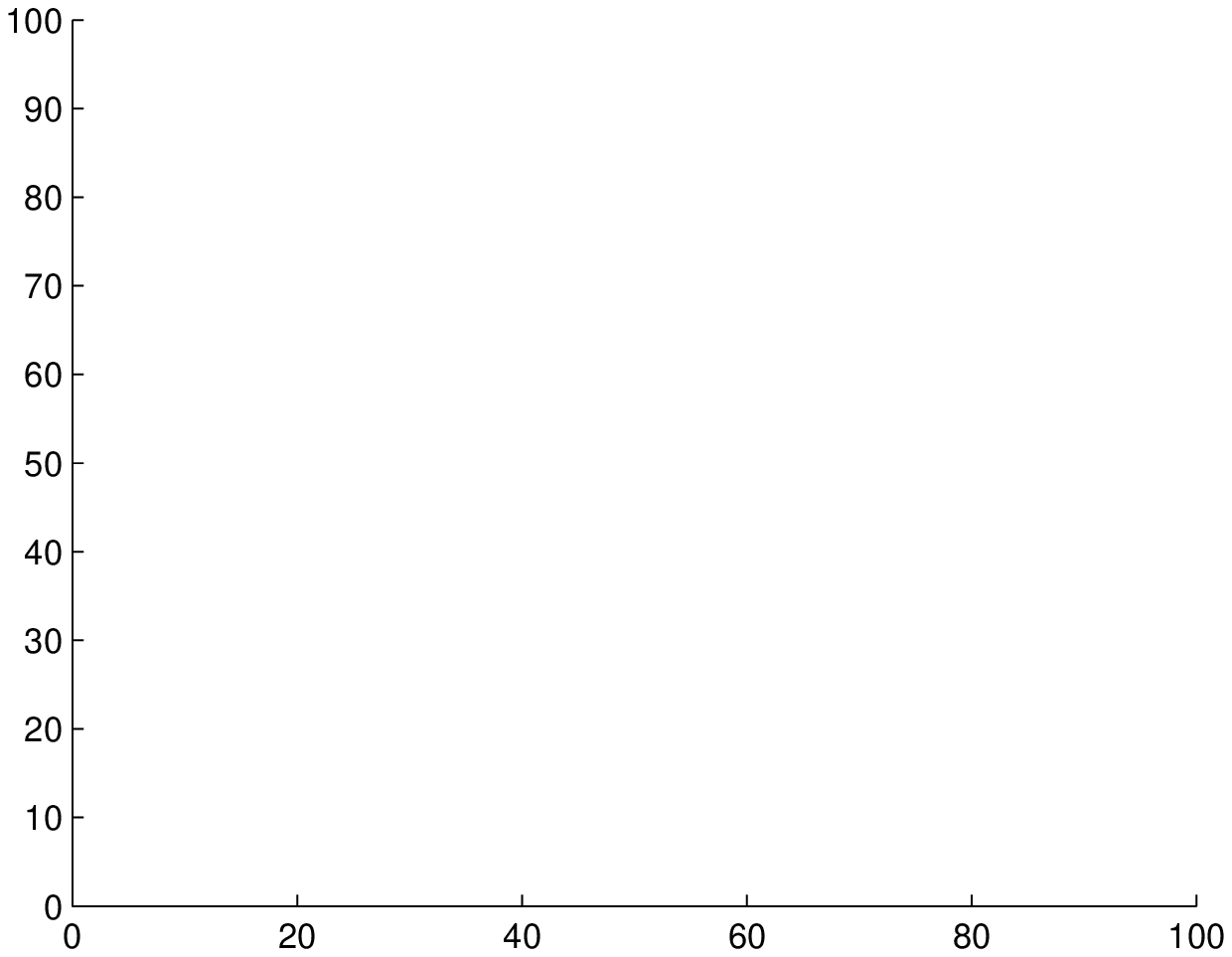}}
\end{center}
\caption{\label{fig_error} Accuracy analysis: the error is small for small $z$, but grows with $z$ and $c$.}
\end{figure}

Figure \ref{fig_error} shows a direct analysis of the accuracy of the variational regular solitary waves obtained above. In this instance, we are able to do a direct accuracy analysis since our variational solution for the regular solitary waves given by \eqref{gtf}, \eqref{solA} and \eqref{solrho} is, unlike for most variational solutions, an analytical one. Inserting this variational solution \eqref{gtf} (with \eqref{solA} and \eqref{solrho}) into the traveling-wave ODE \eqref{SPE_trav}, the deviation of the left-hand side of \eqref{SPE_trav} from zero gives a direct measure of the goodness of the variational solution.

Figure \ref{fig_error} shows this left-hand side for $a_1=1/2, a_3=1$. For all values of the wave-speed $c$, the error is small for small $z$. However, as expected the error increases in the tails of the soliton, i.e., for larger values of $z$, and grows as $c$ increases in magnitude.

\subsection{The variational approximation for embedded solitons}

In the recent and novel variational approach to embedded solitary waves, the
tail of a delocalized soliton is modeled by:
\begin{equation}\label{destail}\phi_{\rm{tail}}=\alpha \cos(\kappa(c)z).\end{equation}
Our embedded solitary wave will be embedded in a sea of such delocalized solitons. The cosine functional form ensures an even solution, and the arbitrary function $\kappa(c)$ will, as shown below, help to ensure the integrability of the action.

Our ansatz for the embedded soliton \cite{es1,es2} uses a second order exponential core model plus the above tail model \eqref{destail}:
\begin{equation}\label{aes}\phi=A \exp\left(-\frac{z^2}{\rho^2}\right)+\phi_{\rm{tail}}.\end{equation}
Plugging this ansatz into the Lagrangian \eqref{anhaL} and reducing the trigonometric powers to double and triple angles yields an equation with trigonometric functions of the double and triple angles, as well as terms linear in $z$. The former would make spatial integration or averaging of the Lagrangian divergent. However, it is possibly to rigorously establish, following a procedure analogous to proofs of Whitham's averaged Lagrangian technique \cite{cb}, that such terms may be averaged out, so we shall set them to zero {\it a priori}.

The terms linear in $z$ would also cause the Lagrangian to be non-integrable. To suppress these, we therefore set:

\begin{equation}\kappa(c)=\pm\frac{\sqrt{- 32 c + 6\alpha^2a_3}}{(\alpha^4 + 32 c^2 - 8 \alpha^2 c)},\end{equation}

which makes linear terms zero. Note that this step, and the preceding step of averaging out trigonometric functions of the higher angles are recent ones for the variational approximation of embedded solitary waves. They are not part of the traditional Rayleigh-Ritz method used for the construction of regular solitary waves.

Next, the rest of the equation can be integrated to give the following action:

\begin{eqnarray}\label{action2}&&\nonumber
\frac{1}{36\rho}\left(\left(\left(\left(\sqrt{3}A^4\frac{27\alpha^4}{4}\left(\frac{1}{2}+\kappa^2\rho^2\right)-18\left(\left(\rho^2\left(c\kappa^2+\frac{3}{2}\right)+c\right)\alpha^2-2c^2\right.\right.\right.\right.\right.\\
&&\left.\left.\left.\left.\left.-2c\rho^2\right)\right)\sqrt{2}
-18\left(\left(-\frac{\alpha^2}{4}(\kappa^2\rho^2+3)+c+\frac{\rho^2}{2}\right)A^2\right)Ae^{\frac{533}{60}\kappa^2\rho^2}\right.\right.\right.\\\nonumber
&&\left.\left.\left.+\frac{108}{125}\left(\left(\frac{125}{8}\rho^2\alpha^4\kappa^2-\frac{250}{3}\left(\frac{3}{4}+c\kappa^2\right)\rho^2\alpha^2+\frac{125}{16}A^3(2+\kappa^2\rho^2)\alpha\right.\right.\right.\right.\right.\\\nonumber
&&\left.\left.\left.\left.+\frac{500}{3}c\rho^2(c\kappa^2+1)\right)e^{\frac{259}{30}\kappa^2\rho^2}
-\frac{125}{3}\left(-\frac{\alpha^2}{8}(1+3\kappa^2\rho^2)+\left(\frac{3}{4}+c\kappa^2\right)\rho^2\right.\right.\right.\right.\\\nonumber
&&\left.\left.\left.\left.+\frac{c}{2}\right)\sqrt{2}A\alpha e^{\frac{503}{60}\kappa^2\rho^2}
-\frac{1000}{81}\left(-\frac{3\alpha^2}{8}\left(3+\frac{5}{2}\kappa^2\rho^2\right)+\left(c\kappa^2+\frac{9}{4}\right)\rho^2\right.\right.\right.\right.\\\nonumber
&&\left.\left.\left.\left.+3c\right)A^2\sqrt{3}e^{\frac{44}{5}\kappa^2\rho^2}
-\frac{250}{3}\left(-\frac{7}{32}\alpha^2\kappa^2+\frac{1}{4}+c\kappa^2\right)\rho^2\alpha^2e^{\frac{199}{30}\kappa^2\rho^2}\right.\right.\right.\\\nonumber
&&\left.\left.\left.+\frac{125}{18}\sqrt{3}\left(\frac{2}{3}+\kappa^2\rho^2\right)A^2\alpha^2e^{\frac{122}{15}\kappa^2\rho^2}
+\frac{125}{16}\sqrt{2}\left(\frac{1}{6}+\kappa^2\rho^2\right)A\alpha^3e^{\frac{413}{60}\kappa^2\rho^2}\right.\right.\right.\\\nonumber
&&\left.\left.\left.+\sqrt{5}\left(\kappa^2\rho^2+\frac{20}{3}\right)A^4e^{\frac{53}{6}\kappa^2\rho^2}+\frac{125}{16}\alpha^4\rho^2\kappa^2e^{\frac{79}{30}\kappa^2\rho^2}\right)\alpha\right)\sqrt{\pi}Ae^{-\frac{533}{60}\kappa^2\rho^2}\right)
\end{eqnarray}


As for the regular solitary waves, the action is now varied with respect to the core amplitude $A$, the core width $\rho$, and the small amplitude $\alpha$ of the oscillating tail.
For strictly embedded solitary waves, which occur on isolated curves in the parameter space where a continuum or ``sea'' of delocalized solitary waves exist, the amplitude of the tail is strictly zero. Once again, this is an extra feature not encountered in the standard variational procedure. Hence, we also need to set $\alpha=0$ in these three variational equations to recover such embedded solitary waves. Implementing this, we have:

\begin{eqnarray}\label{var1}\rho^2\left(12\sqrt{2} c - 6 A^2\right) + 12\sqrt{2} c^2 - 12 c A^2 + \sqrt{6} A^4 &=& 0,\\
\label{var2}\rho^2\left(- 36 \sqrt{2} c + 9 A^2\right) + 36\sqrt{2} c^2 - 18 c A^2 + \sqrt{6} A^4 &=& 0,\\
\label{var3}\frac{-1000\sqrt{3}A^2}{81}\left(\frac{5\rho^2}{4} + 3c\right) \exp\left(\frac{\rho^2}{30c}\right) + \sqrt{5}\left(-\frac{\rho^2}{c} + \frac{20}{3}\right) A^4 &=& 0.      \end{eqnarray}

Subtracting the first two equations \eqref{var1}, \eqref{var2}, one may obtain an expression for $A$ in terms of $\rho$. Solving the two equations obtained by substituting this expression for $A$ into the equation \eqref{var1} and the equation \eqref{var3} yields the solutions
$ c = 0, \rho^2 = 0.2556 c$. Thus, no non-trivial embedded soliton solutions result in this case, i.e. for the SPE equation.

One may also see this from a linearized or tail analysis of the traveling wave equation \eqref{SPE_trav} which does not support oscillatory solutions.

\section{Conclusions}

Three recent analytical approaches have been applied in this paper to treat the possible
classes of traveling wave solutions of a family of so-called short-pulse equations (SPE).

A recent, novel application
of phase-plane analysis is first employed to show the existence of breaking kink wave solutions in certain parameter regimes.

Smooth traveling waves are next considered using a recent technique to derive convergent multi-infinite series solutions for
the homoclinic (heteroclinic) orbits of the traveling-wave equations for the SPE equation, as well as for
its generalized version with arbitrary coefficients. These correspond to pulse (kink
or shock) solutions respectively of the original PDEs. Unlike the majority of unaccelerated convergent series, high
accuracy is attained with relatively few terms. We also show the traveling wave nature of these pulse and front solutions.

Finally, variational methods are employed to treat families of both regular and embedded solitary wave solutions for the SPE PDE.
The technique for obtaining the embedded solitons incorporates several recent generalizations of the usual variational technique and is thus topical in itself. One unusual feature of the solitary waves derived here is that we are able to obtain them in analytical form (within the assumed ansatz for the trial functions). Thus, a direct error analysis is performed, showing the accuracy of the resulting solitary waves.

Given the importance of wave solutions in dynamics and information propagation, and the fact that quite little is known
about solutions of the family of generalized SPE equations considered here, the results obtained are both new and topical.

\end{document}